\documentclass[showpacs,preprint,preprintnumbers,nofootinbib,amsmath,amssymb]{revtex4} 
\usepackage{epsfig}
\let\jnfont=\rm
\def\NPB#1,{{\jnfont  Nucl.\ Phys.\ B }{\bf #1},}
\def\PLB#1,{{\jnfont Phys.\ Lett.\ B }{\bf #1},}
\def\EPJC#1,{{\jnfont Euro.\ Phys.\ J.\ C }{\bf #1},}
\def\PRD#1,{{\jnfont \em Phys.\ Rev.\ D }{\bf #1},}
\def\PRL#1,{{\jnfont Phys.\ Rev.\ Lett.\ }{\bf #1},}
\def\MPLA#1,{{\jnfont Mod.\ Phys.\ Lett.\ A }{\bf #1},}
\def\JPG#1,{{\jnfont J.\ Phys.\ G}{\bf #1},}
\def\CTP#1,{{\jnfont Commun.\ Theor.\ Phys.\ }{\bf #1},}
\def\CPC#1,{{\jnfont Chin. \ Phys. \ C}{\bf #1},}
\def\CPL#1,{{\jnfont Chin. \ Phys. \ Lett}{\bf #1},}
\def\JHEP#1,{{\jnfont JHEP}{\bf #1},}

\def\RMP{\jnfont  Rev. Mod. Phys.}

\def\p_slash{\not{\hbox{\kern-2.1pt $p$}}}
\def\k_slash{\not{\hbox{\kern-2.1pt $k$}}}
\def\E_slash{\not{\hbox{\kern-5.1pt $E$}}}

\newcommand{\be}{\begin{equation}}
\newcommand{\ee}{\end{equation}}
\newcommand{\beq}{\begin{eqnarray}}
\newcommand{\eeq}{\end{eqnarray}}
\newcommand{\bpm}{\begin{pmatrix}}
\newcommand{\epm}{\end{pmatrix}}
\newcommand{\ba}{\begin{array}}
\newcommand{\ea}{\end{array}}

\begin{document}
\title{Charged Higgs Pair Production at the LHC as a Probe of the Top-Seesaw Assisted Technicolor Model}

\preprint{\parbox{1.2in}{\noindent arXiv:1501.01714 }}


\author{Guo-Li Liu$^1$\footnote{guoliliu@zzu.edu.cn}, Xiao-Fei Guo$^1$, Kun Wu$^1$, Ji Jiang$^1$, Ping
Zhou$^{1,2}$ }

\affiliation{ $^1$ Department of Physics, Zhengzhou University, Zhengzhou, 450001, China \\
 $^2$ National Space Science Center, Chinese Academy of Science,100190  }

\begin{abstract}
The top-seesaw assisted technicolor (TC) model, which was proposed recently to explain the 126 GeV
Higgs mass discovered by the Large Hadron Colliders (LHC), predicts light and heavy
charged Higgs bosons in addition to the neutral Higgses. In this paper we will study the
pair productions of the charged Higgs, proceeding through gluon-gluon fusion and
quark-anti-quark annihilation, at the LHC in the frame of the top-seesaw assisted TC model.
We find that in a large part of parameter
space the production cross sections of the light charged Higgs pair at the LHC can be quite
large compared with the low standard model backgrounds, while it is impossible for the pair
production of the heavy ones to be detected with the strong final mass suppression.
Therefore, at the LHC future experiments, the light charged Higgs
pair production may be served as a probe of this new TC model.
\end{abstract}

\maketitle

\section{Introduction}
Though it is successfully tested by various high energy experiments,
including the 126 GeV Higgs \cite{7.4} discovery by the Large Hadron Collider (LHC) in CERN \cite{lhc},
the standard model (SM) of particle physics \cite{sm}
is still believed by many people to be an effective theory below certain high energy scale.
The origin of the mechanism for electroweak symmetry breaking(EWSB) as well as the
Yukawa couplings remain a mystery in current particle physics.
Besides, the neutrino oscillation experiments indicate that neutrinos are massive,
which manifestly requires new physics beyond the standard model
\cite{neutrino}. At the same time, SM itself cannot provide viable dark matter candidates \cite{dark-matter}.
Therefore, it is interesting from both the theoretical point of view and the experimental
search aspects to extend the standard model to understand the EWSB mechanism and possibly extended the Higgs sector.

The TC-type models\cite{technicolor,tc2-rev}, in which EWSB can be achieved via introducing the new
strong interaction-- the TC interaction, without the aid of the elementary scalar Higgs
boson \cite{9712319,9809470,9908391,9911266,he-seesaw,1304.2257}, could completely avoid the problems
 arising from the elementary Higgs scalar field in the SM. The TC
models open up new possibilities for new physics beyond the SM,
and might produce observed signatures in future high energy
collider experiments.

Among various kinds of TC theories, the topcolor
scenario\cite{topcolor} is attractive because it can not only provide a possible
dynamical EWSB mechanism, but explain the large top quark mass simultaneously.
These traditional TC theories, however,
have encountered a severe obstruction since they are difficult
to provide a light scalar candidate. To solve the problem, top-seesaw assisted TC
model\cite{he-seesaw,top-seesaw-tc} is proposed, which
requires EWSB are shared between different contributions, i.e. there exists different scalars,
with different value expectation values (VEVs), say $v_1$ and $v_2$, satisfying
$v^2_{EW}= v^2_1+ v^2_2$, with $v_1(v_2)< v_{EW}$, the electroweak scale.
Then the masses of the excitations in different sections, which are dictated by $v_1$ and $v_2$,
may also be smaller than $v_{EW}$.

With the enlarged gauge group, the top-seesaw assisted TC model predicts
more Higgs bosons, including the additional charged scalars. Actually, the existence
of new charged scalars are predicted in many new physics theories,
such as the supersymmetry\cite{susy}, TC (topcolor)\cite{technicolor,tc2-rev,topcolor},
little Higgs\cite{lh-review} and the left-right twin Higgs\cite{lrth-review}, etc.
These charged new scalars may have very large signals at the colliders,
and If we can find any evidence of them, it would necessarily be the signal of the
new physics beyond the SM. Thus, studying the signals of the charged scalars\cite{char-higgs}
at the running LHC will be of special interest.

As we know, the pair productions of the charged scalars, at the tree-level or the one-loop level,
may have very large production rates \cite{pair-production},
so in the top-seesaw assisted TC model, we can consider the pair
production of the new charged scalars at the LHC, and analysis the observable possibility, which
may serve as a good channel to probe such new TC model.

In this paper, we will study how the top-seesaw assisted TC model constrains the scalar
pair production processes $gg\to S^+S^-$ and $q\bar q \to S^+S^-$ ($S^\pm$ denotes the charged
scalars and $q=u,d,c,s,b$ quarks).
We will calculate the cross sections of these processes and compare the signals with their SM
backgrounds.

In Sec.~II, the newly proposed TC model relative to our calculations is
reviewed and the new couplings related to the scalar pair production processes
$gg\to S^+S^-$ and $q\bar q\to S^+S^-$($q=u,d,c,s,b$ quarks) at the LHC are also
given in this section. Sec.~III shows the numerical results of these processes and
analysis simply the SM backgrounds and the detectable probability of the final state at the the LHC.
Our summary and discussions are given in Sec.~IV.
\section{The top-seesaw assisted TC model and the relevant couplings}
To solve the phenomenological difficulties of traditional
TC theory, the top-seesaw assisted TC model\cite{top-seesaw-tc} was proposed by
adding new vector-like quarks in the TC models. The basic idea of the models is to
combine top-seesaw model\cite{9712319,9809470,9911266,he-seesaw,1304.2257} with TC
model\cite{technicolor} in a way similar to topcolor assisted TC (TC2)\cite{tc2-rev} models.
In this new model, masses of all leptons and the light
quarks are assumed to be generated by some underlying ETC dynamics
operating at much higher scales and the mass patterns of the third and fourth quark generations
are mainly provided dynamically by the seesaw mechanism.

\subsection{The low energy effective lagrangian of the top-seesaw assisted model}
The underlying gauge symmetry in the ultraviolet (UV) part of
top-seesaw theory is $SU(3)_1\times SU(3)_2\times SU(2)_L
\times U(1)_1 \times U(1)_2$, which is broken to $SU(3)_{QCD}\times SU(2)_L
\times U(1)_Y$, generating $8 + 1$ massive gauge bosons $G'$ and $Z'$, which masses are
assumed at the same order, denoted as $M_V$.
At low energies, the interactions via the $8 + 1$ massive gauge bosons
exchange lead to effective four fermion interactions, of which the terms that interest us are
given as
\beq
{\cal L}_S^{4f} =
G_b \left( \bar{D}^{(4)}_R Q^{(3)}_L\right)^2
+
G_t \left( \bar{U}^{(4)}_R Q^{(3)}_L\right)^2
+
G_{tb} \left( \bar{Q}^{(3)}_L U^{(4)}_R \right)
\left( \bar{D}^{(4) c}_R i \tau_2 Q^{(3) c}_L.\right)+h.c
\,,
\label{4f-present}
\eeq
where $G_{t,b} $ are the scalar mass terms and $G_{tb}$ are the diagonal terms and we here will not
discuss them in detail, since every coupling that we will obtain is actually closely related
to the specific form of different fields, which will be discussed later.

In this section, we will consider the low energy effective
Lagrangian for the four fermion interaction sector and its mixing with the TC
sector, of which, the dynamical top seesaw sector based on the conventional
Nambu-Jona-Lasinio (NJL) model \cite{njl}, can be given by the fermion bubble
sum approximation \cite{he-seesaw,bardeen},
The low energy effective Lagrangian valid for $\mu <\Lambda \simeq M_G' \simeq M_Z'$,
where $\mu$ is the scale of the theory after the gauge breaking $ SU(3)_1\times SU(3)_2 \times U(1)_1 \times U(1)_2
 \to SU(3)_{\rm QCD} \times U(1)_Y$ and $\Lambda$ is ultraviolet (UV) cutoff.

The auxiliary Higgs fields $\Phi_{1,2}$ are introduced with
$\Phi_1\sim \bar D_R^{(4)}Q_L^{(3)}$ and $\Phi_2\sim \bar U_R^{(4)} Q_L^{(3)}$,
and $\Phi_{1,2}$ ($i=1,2$) can be further parameterized as,
\beq
\Phi_i = \left( \ba {c} \pi^+_i \\[1ex] \dfrac{1}{\sqrt{2}} [ v_i + h^0_i - i \pi^0_i] \\ \ea \right).
\label{higgs}
\eeq

As we know, the top-seesaw assisted TC model includes two sections, where one
sector, i.e, the top seesaw section, generates the large top quark mass
and partially contributes to EWSB while the other sector, i.e,
TC interaction, is responsible for the bulk of EWSB and the
generation of light fermion masses.
The Nambu-Goldstone bosons (NGBs) in the TC sector can be described as
the most minimal electroweak chiral Lagrangian\cite{ewcl-appelquist-longhitano}
according to the the most minimal structure breaking $G/H = [SU(2)_L\times SU(2)_R/SU(2)_V $,
 so the leading order chiral Lagrangian can be written as
\beq
{\cal L}^{(2)}_{\rm EWCL}
=
\left| D_\mu \Phi_{\rm TC} \right|^2\,, ~~ with~~~~~~~ \Phi_{\rm TC} = \bpm \pi^+_{\rm TC} \\[2ex] \dfrac{1}{\sqrt{2}}\left[ v_{\rm TC} - i \pi^0_{\rm TC}\right] \epm\, ,
\label{chiral-Lag-kin}
\eeq
where $\tilde{\Phi}_{\rm TC} \equiv i \tau^2 \Phi^*_{\rm TC}$. 
The covariant derivative $D_\mu \Phi_{\rm TC}$ is
\beq
D_\mu \Phi_{\rm TC} = \partial_\mu \Phi_{\rm TC} -  i  g W^a_\mu T^a \Phi_{\rm TC} - \frac{1}{2} g' B_\mu \Phi_{\rm TC}\,,
\label{higgs-covariant-derivative}
\eeq
where $T^a = (1/2) \tau^a$, and  $g$ and $g'$ are gauge couplings of the SM $SU(2)_L,U(1)_Y$
gauge boson fields $W_\mu,B_\mu$, respectively.

The reason for the missing CP-even component of the $\Phi_{\rm TC}$ in Eq.(\ref{chiral-Lag-kin}), is that,
 the Higgs effects are found to be small \cite{ewcl-appelquist-longhitano},
since the Nambu-Goldstone bosons (NGBs) in the TC sector, which are described
by the most minimal structure of the electroweak chiral Lagrangian\cite{ewcl-appelquist-longhitano},
can be a strongly interacting heavy-Higgs-boson sector, i.e.,
the gauged nonlinear $\sigma$ model, i.e., the nonrenormalizability of the no-Higgs-boson theory.
And furthermore, we have also assumed that the TC section only provides the very small
masses of the light fermions in a higher scale, so the effects of the "Higgs" from TC sector
at low energy are negligible, compared to those of the top-seesaw sector. Actually, in this model we will set
$m_{ETC} = \Lambda_{TC} = 4\pi v_{TC}$ corresponding to the cutoff scale for the non-linear sigma model which we use to describe the TC sector \cite{top-seesaw-tc}.

At the low energy, the effective Lagrangian concerning Higgs section in the top-seesaw assisted TC model
can be explicitly written by
\beq
{\cal L}_{\rm higgs} (\Phi_1,\Phi_2,\Phi_{\rm TC})
=
\sum_{i=1,2,{\rm TC}} \left| D_\mu \Phi_i \right|^2
+ {\cal L}_{\rm yukawa}
- V(\Phi_1,\Phi_2,\Phi_{\rm TC}) \,.
\label{hybrid-full-EFT}
\eeq
where the covariant derivatives of $\Phi_i$ are the same forms as that in Eq.(\ref{higgs-covariant-derivative}) and the effective Yukawa interaction terms ${\cal L}_{\rm yukawa}$ are
 \beq
{\cal L}_{\rm yukawa}
\!\!&=&\!\!
- \!\!\!\sum^{\text{quarks}}_{i,j=1,2,3} y^{(d)}_{ij} \bar{Q}^{(i)}_L \Phi_{\rm TC} D^{(j)}_R
- \!\!\!\sum^{\text{quarks}}_{i,j=1,2,3} y^{(u)}_{ij} \bar{Q}^{(i)}_L \tilde{\Phi}_{\rm TC} U^{(j)}_R
\nonumber\\[1ex]
&&
-  y_1 \bar{Q}^{(3)}_L \Phi_1 D^{(4)}_R
- y_2 \bar{Q}^{(3)}_L \tilde{\Phi}_2 U^{(4)}_R + {\rm h.c., }
\label{reno-yukawa-H4G}
\eeq
where the Yukawa couplings $y_{1,2}$  and $y_{ij}$ in the above
equation are given later, when discussing the Yukawa terms of the $3,4$ generations.

The potential $V(\Phi_1,\Phi_2,\Phi_{\rm TC})$ in Eq.(\ref{hybrid-full-EFT}) can be defined as
two sections
\beq
V(\Phi_1,\Phi_2,\Phi_{\rm TC})
=
V_{\rm TSS}(\Phi_1,\Phi_2)
+
V_M(\Phi_1,\Phi_2,\Phi_{\rm TC})
\label{3HDM-potential}
\,.
\eeq
Similar to the attainment of the Yukawa terms in Eq. (\ref{reno-yukawa-H4G}), the former part of the above Higgs potential can be given as,
\beq
V_{\rm TSS}(\Phi_1,\Phi_2)
&=&
M^2_{11} |\Phi_1|^2 + M^2_{22} |\Phi_2|^2
- M^2_{12} \left[ \Phi^\dagger_1 \Phi_2 + {\rm h.c.}\right]
\nonumber\\[1ex]
&&
+ \frac{1}{2} \lambda_1( \Phi^\dagger_1 \Phi_1 )^2
+ \frac{1}{2} \lambda_2 ( \Phi^\dagger_2 \Phi_2)^2
+ \lambda_3(\Phi^\dagger_1 \Phi_1)(\Phi^\dagger_2 \Phi_2)
+ \lambda_4(\Phi^\dagger_1 \Phi_2)(\Phi^\dagger_2 \Phi_1)
\,,
\label{TSS-higgs-potential}
\eeq
where $M_{ij}^2~(i,j=1,2)$ are the Higgs mass terms and $\lambda_{1,2,3,4}$, Higgs quartic couplings.
$M_{ij}^2~(i,j=1,2)$ can be confined by the scalar masses, while $\lambda_{1,2,3,4}$ can be constrained
by solving the RGEs with the compositeness conditions\cite{bardeen,compo-con-hashimoto} of this
 model, and we take $\lambda_1=\lambda_1=\lambda_1=\lambda_1=1$ in this paper, since they are in the order
 of ${\cal O}(1)$ \cite{top-seesaw-tc}.

Different from the obtainment of the Yukawa terms and the potential $V_{TSS}(\Phi_1,\Phi_2)$, which are both arising
from the underlying theory of the four fermion interactions in Eq.(\ref{4f-present}),
the terms $V_M(\Phi_1,\Phi_2,\Phi_{\rm TC})$, which are the mixing between the TC sector
and the top-seesaw sector \cite{Chivukula:2011ag}, can be written as
\beq
V_M(\Phi_1,\Phi_2,\Phi_{\rm TC})
\!\!\!&=&\!\!\!
c_1 v^2_1  \left| \Phi_1- \frac{v_1}{v_{\rm TC}}\Phi_{\rm TC} \right|^2
+ c_2 v^2_2 \left| \Phi_2- \frac{v_2}{v_{\rm TC}}\Phi_{\rm TC} \right|^2\,,
\label{mixing-TSS-MWT}
\eeq
where $c_{1,2}$ are dimensionless parameters of $\cal O (1)$ and we will take $c_1=c_2=1$ in our calculations.

 Under the above definitive scalars, we know that the vacuum structure of this model 
 is determined by three vacuum expectation values (VEVs)
of the three scalar doublets,  $v_{{\rm TC},1,2}$, which all contribute to EWSB and satisfy the
relation $v_1^2+v_2^2+v_{\rm TC}^2=v_{\rm EW}^2$ with
 $v_{\rm EW}=246~ GeV$.  Mixing angles $\beta$ and $\phi$ are introduced with the definition as
\beq
\tan \beta \equiv \frac{v_2}{v_1},
\quad
\tan^2 \phi \equiv \frac{v^2_{\rm TC}}{v^2_1 + v^2_2}\, ,
\eeq
or $
 v_{\rm TC} = v_{\rm EW} \sin \phi, \hspace{1cm} v_1 = v_{\rm EW} \cos \phi \cos \beta, \hspace{1cm} v_2 = v_{\rm EW} \cos \phi \sin \beta $.

\subsection{The Higgs boson spectrum in the present model}
From the scalar doublets shown in Eq.(\ref{higgs}) and Eq.(\ref{chiral-Lag-kin}),
we know that there should be $11$ scalars, three of which, however,
will become the longitudinal components of the electroweak bosons, in the proper
parameterization form, so there should be $8$ scalars left. Since $1$ CP-odd neutral
and $2$ charged bosons will be "eaten", there should exist $2$ CP-odd, $2$ CP-even, $4$
charged Higgs. In the following, we will consider the mixing and coupling with the other
particles concerned in this paper.

We can write down the quadratic terms of the NGB fields via the potentials $V_{TSS}(\Phi_1,\Phi_2)$ and $V_M(\Phi_1,\Phi_2,\Phi_{TC})$ in Eq.(\ref{TSS-higgs-potential}) and Eq.(\ref{mixing-TSS-MWT}) as
\beq
{\cal L}^{\rm qd} =
-\frac{1}{2}(\pi^0_1 \,\,\, \pi^0_2 \,\,\, \pi^0_{\rm TC}) {\cal M}^2_{\pi} \bpm \pi^0_1 \\ \pi^0_2 \\ \pi^0_{\rm TC}\epm
-(\pi^+_1 \,\,\, \pi^+_2\,\,\,\pi^+_{\rm TC}) {\cal M}^2_{\pi\pm} \bpm \pi^-_1 \\ \pi^-_2 \\\pi^-_{\rm TC}\epm
-\frac{1}{2}(h^0_1 \,\,\, h^0_2) {\cal M}^2_{h} \bpm h^0_1 \\ h^0_2 \epm .
\label{lag-higgs-mass}
\eeq
The mass matrix of the charged Higgs sector is,
\beq
\left. {\cal M}^2_{\pi \pm} \right|_{\rm TC = 0} =
\left[ M^2_{12} - \frac{1}{2}  \lambda_4 v^2_{\rm EW} \cos^2 \phi \sin \beta \cos \beta\right]
\bpm \tan \beta & -1 \\ -1 & \tan \beta \epm
\,,\label{TSS-charged-higgs-mass}
\eeq
where $M^2_{12}$ can be treated as the free parameters.

Due to the mixing of the top-seesaw and TC sectors,
the mass matrix of the charged CP-odd Higgs boson fields, $\pi^\pm_i(i=1,2,{\rm TC})$, can
be given as
\beq
{\cal M}^2_{\pi \pm} =
\left(
\begin{array}{cc|c}
&\mbox{\raisebox{-2ex}{\large$\left. {\cal M}^2_{\pi \pm} \right|_{\rm TC = 0}$}}&0\\
&&0\\ \hline
0&0&0
\end{array}
\right)
+\bpm
 c_1 v^2_1& 0& - M^2_1
\\[1ex]
0&  c_2 v^2_2 & - M^2_2
\\[1ex]
- M^2_1   & - M^2_2   & M^2_1 \cos \beta \cot \phi + M^2_2 \sin \beta \cot \phi
\epm\,.
\label{mass-CPNGB}
\eeq
where $c_1(c_2)$ is a dimensionless parameter and $M^2_1 = c_1 v^2_1 \frac{v_1}{v_{\rm TC}}$, $M^2_2 = c_2 v^2_2 \frac{v_2}{v_{\rm TC}}$.

In terms of the mass basis, the CP-odd neutral Higgs bosons and the charged Higgs bosons can be given as
\beq
\bpm G^0 \\[1ex] A^0_2 \\[1ex] A^0_1 \epm =
O^T_0
\bpm \pi^0_1 \\[1ex] \pi^0_2\\[1ex] \pi^0_{\rm TC} \epm
\quad , \quad
\bpm G^\pm\\[1ex] H^\pm_2 \\[1ex] H^\pm_1 \epm =
{\cal O}^T_\pm
\bpm \pi^\pm_1 \\[1ex] \pi^\pm_2\\[1ex] \pi^\pm_{\rm TC} \epm \,,
\label{mixing-mass-inter}
\eeq
with the orthogonal matrix $O_p\,\,(p=0,\pm)$ as \cite{Hashimoto:2009ty}
\beq
O_p
= \bpm
\cos \phi \cos \beta & - \sin \beta \cos \zeta_p + \sin \phi \cos \beta \sin \zeta_p &  -\sin \beta \sin \zeta_p - \sin \phi \cos \beta \cos \zeta_p
\\[1ex]
\cos \phi \sin \beta & \cos \beta \cos \zeta_p + \sin \phi \sin \beta \sin \zeta_p & \cos \beta \sin \zeta_p - \sin \phi \sin \beta \cos \zeta_p
\\[1ex]
\sin \phi & -\cos \phi  \sin \zeta_p & \cos \phi \cos \zeta_p
\epm
\label{ortomatrix-PNGB}
\,.
\eeq
Here the mixing angle between the mass and interaction eigenstates $\tan \zeta_p$ is composed as
\beq
\tan \zeta_p =
\frac{\hat{M}^2_{S_2} \cos \phi  \sin \phi- \left( M^2_1 \cos \beta+ M^2_2 \sin \beta \right) }{\sin \phi \left( M^2_1 \sin\beta -  M^2_2 \cos \beta \right) }\,.
\eeq
\subsection{The couplings of the charged Higgs boson to the third and the fourth generation quarks}
We will discuss the mixing between the third generation quarks and their vector-like
partners, i.e., the fourth quarks. Firstly we find the fermion
mass part after the dynamical EWSB,
\beq - \bpm \bar{U}^{(3)}_L & \bar{U}^{(4)}_L\epm
\bpm 0 & \Sigma_U \\[1ex] M^{(43)}_{U} & M^{(44)}_U\epm
\bpm U^{(3)}_R \\[1ex] U^{(4)}_R\epm
- \bpm \bar{D}^{(3)}_L & \bar{D}^{(4)}_L\epm
\bpm 0 & \Sigma_D \\[1ex] M^{(43)}_{D} & M^{(44)}_D\epm
\bpm D^{(3)}_R \\[1ex] D^{(4)}_R\epm
+ {\rm h.c.}   \,,
\label{seesaw-mass-part}
\eeq

Now, the quark mixing matrices $U, D$ was presented to reflect the seesaw mechanism
for the third and the fourth generation, and the quark
mixing matrices are given as \cite{top-seesaw-tc}
\beq
U^L_{\alpha\beta}\simeq
\bpm
1 & 0 & 0 & 0 \\
0 & 1 & 0 & 0 \\
0 & 0 & c^t_L & s^t_L \\
0 & 0 & -s^t_L & c^t_L
\epm
\quad &,& \quad
U^R_{\alpha\beta}
\simeq
\bpm
1 & 0 & 0 & 0 \\
0 & 1 & 0 & 0 \\
0 & 0 & -c^t_R & s^t_R \\
0 & 0 & s^t_R & c^t_R
\epm\,,
\,\label{seesaw-mixing-u}
\\[2ex]
D^L_{\alpha\beta} =U^L_{\alpha\beta}\left.\right|_{t \to b}
\quad \hspace*{8ex} &,& \quad
D^R_{\alpha\beta} = U^R_{\alpha\beta}\left.\right|_{t \to b}\,.
\label{seesaw-mixing-d}
\eeq
where $c^t_L \equiv \cos \theta^t_L\,,\,s^t_R \equiv \sin \theta^t_R$, etc. These fermion mixing
matrices $U$ and $D$ in the above two equations diagonalize the
mass mixing matrices in Eq.(\ref{seesaw-mass-part}),
and the eigenvalues of them are  $m_{t,b}(TSS)$ (masses of top and bottom quarks generating by top-seesaw),
and $m_{T,B} $(mass of the vector-like partner of the third generation quark), with $m_{T,B} > m_{t,b}(TSS)$.
So $c^{t,b}_L,s^{t,b}_R$ can be written as
\beq
&&
[c^t_L]^2 \equiv \frac{m^2_T - \Sigma^2_U}{m^2_T - m^2_t(\text{TSS})}, \quad
[s^t_R]^2 \equiv \frac{m^2_t(\text{TSS})}{\Sigma^2_U} [c^t_L]^2
\,,\label{mixing-top}
\\[1ex]
&&
[c^b_L]^2 \equiv \frac{m^2_B - \Sigma^2_D}{m^2_B - m^2_b(\text{TSS})}, \quad
[s^b_R]^2 \equiv \frac{m^2_b(\text{TSS})}{\Sigma^2_D} [c^b_L]^2
\,.\label{mixing-bottom}
\eeq

In this model, the Yukawa terms for third generation quarks and their vector-like
partners, i.e., the fourth quarks, which is a part of Eq.(\ref{reno-yukawa-H4G}),
are written explicitly as
\beq
{\cal L}^{\rm 3-4}_{\rm yukawa}
= - y_1 \bar{Q}^{(3)}_L \Phi_1 D^{(4)}_R - y_2 \bar{Q}^{(3)}_L \tilde{\Phi}_2 U^{(4)}_R
- y^b_{\rm TC} \bar{q}_L \Phi_{\rm TC} b_R - y^t_{\rm TC} \bar{q}_L \tilde{\Phi}_{\rm TC} t_R
+ \text{h.c.}
\,,\label{3higgs-yukawa}
\eeq
where the couplings $y^{1,2}$ and $y^{b,t}_{\rm TC}$ are given by\cite{top-seesaw-tc}
\beq
y_1 = \frac{\sqrt{2} \Sigma_D}{v_1}, \quad
y_2 = \frac{\sqrt{2} \Sigma_U}{v_2},\quad
y^b_{\rm TC} = \frac{\sqrt{2} \epsilon_b m_b}{v_{\rm TC}}, \quad
y^t_{\rm TC} = \frac{\sqrt{2} \epsilon_t m_t}{v_{\rm TC}},
\eeq
Note that $y_{1,2}$ is obtained via the renormalization group
equations (RGEs) and according to the discussion in Ref. \cite{top-seesaw-tc},
$y_1=y_2=2$ is appropriate. From the definitions of $y^{b,t}_{\rm TC}$ and the
couplings in Eq.(\ref{3higgs-yukawa}) we can see clearly that
the parameters $\epsilon_t$ and $\epsilon_b$ are the fraction
of the ETC interactions to the masses of the top and bottom quarks, respectively.
In order to realize the top-seesaw dynamics, we must
have $\Sigma_U >m_t(TSS)=(1-\epsilon_t)m_t$ with $\epsilon_t<0.1$, so we can take $\Sigma_U$
as a free parameter only if the seesaw condition mentioned above is satisfied.

Taking the Eq.(\ref{mixing-mass-inter}) and Eq.(\ref{seesaw-mixing-u})
into Eq.(\ref{3higgs-yukawa}), by which the charged Higgs and the quarks are
changed into mass eigenstates, we can obtain the couplings of the charged Higgs
to the heavy quarks,
\beq \nonumber
{\cal{L}}_{Hff'}& =&  (a^L_{tb}+b^L_{tb}\gamma^5)  H^+_L \bar{t} b  +  (a^H_{tb}+b^H_{tb}\gamma^5) H^+_H \bar{t} b
+ (a^L_{Tb}+b^L_{Tb}\gamma^5)  H^+_L \bar{T} b + (a^H_{Tb}+b^H_{Tb}\gamma^5)  H^+_H \bar{T} b \\ \nonumber &&
+ (a^L_{tB}+b^L_{tB}\gamma^5)  H^+_L \bar{t} B  +  (a^H_{tB}+b^H_{tB}\gamma^5) H^+_H \bar{t} B
+ (a^L_{TB}+b^L_{TB}\gamma^5)  H^+_L \bar{T} B \\   && +  (a^H_{TB}+b^H_{TB}\gamma^5) H^+_H \bar{T} B
+h.c.
\label{hff-couplings}
\eeq
where
\beq
 a^L_{tb} &=&  \frac{1}{2}(-y_1 c_L^ts_R^b O_{21}+ y_2 s^t_R c^b_L O_{22}-y_{TC}^bO_{23}+ y^t_{TC}O_{23} ),~ \\
a^L_{tB} &=&  \frac{1}{2}(- y_1 c^t_L c^b_R O_{21} - y_2 s^b_L s^t_R O_{22}),
\\ 
 a^L_{Tb} &=&  \frac{1}{2}( y_1 s_L^ts_R^b O_{21}+ y_2 c^t_R c^b_L O_{22}),
\\ 
 a^L_{TB} &=&  \frac{1}{2}( y_1 s_L^tc_R^b O_{21}- y_2 c^t_R s^b_L O_{22}),
\\ 
b^L_{q_iq_j} &=&  a^L_{q_iq_j} |_{y_2->-y_2, y^t_{TC}->-y^t_{TC}}; ~~ a^H_{q_iq_j}(b^H_{q_iq_j}) =  a^L_{q_iq_j}(b^L_{q_iq_j})|_{O_{2i}->O_{3i}};
\eeq
with $q_i,q_j=t,b,T,B$ ($i\neq j$) quarks.

\subsection{The couplings of the charged Higgs boson pair to neutral Higgs}

Diagonalizing the fermions and scalars in the the Higgs mixing potential $V_M$ in Eq.(\ref{hybrid-full-EFT}), which are related to three scalars couplings, we can arrive in the three scalars couplings as,
\be
{\cal{L}}_{scc} =  Y^L_{hcc}  h^0H^+_L H^-_L  +  Y^L_{Hcc}  H^0H^+_L H^-_L  +Y^H_{hcc}  h^0H^+_H H^-_H
+Y^H_{Hcc}  H^0H^+_H H^-_H
\label{hchch-couplings}
\ee
Where
\beq \nonumber
 Y^L_{hcc} &=&  -\lambda_1 v_1 s_\alpha (O_p^-)_{13}^2  +\lambda_2 v_2 c_\alpha (O_p^-)_{23}^2-\lambda_3v_1 s_\alpha (O_p^-)_{23}^2+\lambda_3v_2 c_\alpha (O_p^-)_{13}^2 +\lambda_4 c_h(O_p^-)_{23}(O_p^-)_{13},\\  \nonumber
 Y^L_{Hcc} &=&   \lambda_1 v_1 c_\alpha (O_p^-)_{13}^2  +\lambda_2 v_2 s_\alpha (O_p^-)_{23}^2+\lambda_3v_1 c_\alpha (O_p^-)_{23}^2+\lambda_3v_2 s_\alpha (O_p^-)_{13}^2 +\lambda_4 c_H(O_p^-)_{23}(O_p^-)_{13},\\  \nonumber
 Y^H_{hcc} &=&   -\lambda_1 v_1 s_\alpha (O_p^-)_{12}^2  +\lambda_2 v_2 c_\alpha (O_p^-)_{22}^2-\lambda_3v_1 s_\alpha (O_p^-)_{22}^2+\lambda_3v_2 c_\alpha (O_p^-)_{12}^2 +\lambda_4  c_h(O_p^-)_{23}(O_p^-)_{12},\\  \nonumber
 Y^H_{Hcc} &=&   \lambda_1 v_1 c_\alpha (O_p^-)_{12}^2  +\lambda_2 v_2 s_\alpha (O_p^-)_{22}^2+\lambda_3v_1 c_\alpha (O_p^-)_{22}^2+\lambda_3v_2 s_\alpha (O_p^-)_{12}^2 +\lambda_4 c_H (O_p^-)_{23}(O_p^-)_{12}. \\ 
\eeq
Here $s_\alpha=\sin\alpha$, $c_\alpha=\cos\alpha$, and $\alpha$ is the neutral Higgs mixing, with
\beq
\bpm H^0 \\ h^0 \epm  =
\bpm \cos \alpha & \sin \alpha \\ -\sin \alpha & \cos \alpha \epm
\bpm h^0_1 \\ h^0_2 \epm
.\label{CP-even-higgs-matrix}
\eeq
The matrix $O^{-}_p$ is the inverse of the psedu-Goldstone boson mixing matrix of $O_p$ shown in Eq.(\ref{ortomatrix-PNGB}),
 which can be given as
\beq
O^-_p = \bpm
\cos\phi \cos\beta & -\cos\zeta_p \sin\beta +
      \cos\beta \sin\phi \sin\zeta_p &  -\cos\beta \cos\zeta_p \sin\phi -
      \sin\beta \sin\zeta_p \\ 
\cos\phi \sin\beta &  \cos\beta \cos\zeta_p + \sin\phi \sin\beta \sin\zeta_p & -\cos\zeta_p \sin\phi \sin\beta + \cos\beta \sin\zeta_p \\ 
\sin\phi& -\cos\phi \sin\zeta_p& \cos\phi \cos\zeta_p
\label{inverse-ortomatrix-PNGB}\epm.
\eeq
The coupling constants $c_h$ and $c_H$ are written as,
\beq c_H=v_1 s_\alpha+v_2 c_\alpha, ~~~c_h=v_1 c_\alpha-v_2 s_\alpha
\eeq

\subsection{The relevant couplings $\gamma \bar b b$ and $Z^0 \bar b b$}
The other relevant couplings are the gauge bosons with the charged Higgs pair and the
 $V \bar b b$ ($V=\gamma,Z^\mu$) interactions, the former of which are the same as
 those in SM and the latter are given as\cite{top-seesaw-tc},
\be
 \gamma^\mu \bar b b: ieQ_b\gamma^\mu,~~~~~
Z^\mu \bar b b: \frac{ig}{c_W}[(g_L^b +\frac{1}{4}(s_L^b)^2) -\frac{1}{4}(s_L^b)^2\gamma^5 ]
\ee
\subsection{The simple discussions of the relevant model parameters }
Obtaining the relevant couplings, we will now discuss the parameters involved in the models.
The parameters of this models related to our discussions are $c_1$, $c_2$, the mixing angles $\beta$, $\phi$
the scale $\Lambda$, the ETC contributions to the masses of the top and the bottom
quarks $\epsilon_t$ and $\epsilon_b$, the vector-like quark mass $m_T$ and $m_B$, the scalar masses
and the three VEVs of the doublets $v_1,~v_2,~v_{TC}$, which satisfy
$v_1^2+v_2^2+v_{TC}^2=v_{EW}^2$. 
The vector-like quark mass $m_T$ and $m_B$ are constrained
by the oblique parameter and can be chosen as $m_T=m_B=5$ TeV\cite{top-seesaw-tc}.
The scalar section, the lighter CP-even Higgs are chosen to be the 126 GeV
SM-like Higgs, and the charged ones are thought to be heavier than that \cite{top-seesaw-tc}.  
Now we simply discuss the constraints of the relevant parameters.
\begin{itemize}
\item[{\rm (1)}]
 The compositeness scale $\Lambda$ is identified with the mass scale of
the massive coloron $M_G'$ in the present models.  With the constraints of the
$M'_G$ \cite{top-seesaw-tc}, 
 $\Lambda>4$ TeV, a light CP-even Higgs with mass around 126 GeV can be accommodated within the model for
arbitrary  $\Lambda$ with $4< \Lambda<100$ TeV. To diminish the contributions from the massive topcolor gauge bosons
to the electroweak precision parameters at the same time, we here take $\Lambda=50$ TeV \cite{top-seesaw-tc}.

\item[{\rm (2)}] The couplings $y_{1,2}$ are solved from RGEs and the compositeness conditions \cite{bardeen,compo-con-hashimoto}.
From 
 Ref. \cite{top-seesaw-tc}, we can see that if the $\Lambda$
is assumed to be at about $50$ Tev, it is suitable to
take $y_1=y_2=2$, which will be applied in our discussion.

\item[{\rm (3)}]  About the $\epsilon_t$ and $\epsilon_b$ parameter, which are the
fraction of the ETC contribution to the top and bottom quark masses. Generally,
we take this parameter small, $0 < \epsilon_{t,b} <0.1$, which means that the
ETC contribution to the heavy mass is smaller than that of the
seesaw section, i.e., the heavy fermion masses are mainly provided by the seesaw mechanism.

\item[{\rm (4)}] About the mass bounds for the vector-like quarks,
in order that fermion sector does not generate a large contribution to the T-parameter,
the masses can be set as $m_T =m_B= 5$ TeV \cite{top-seesaw-tc}.

\item[{\rm (5)}] About the mass bounds for the charged Higgs, of which we will consider the
pair production at the LHC, we assume the light ones with a mass larger than
$200$ GeV, while masses of the heavy ones are in the range from 1000 to 5000 GeV.

\item[{\rm (6) }] About the mixing angles $\beta$, $\phi$, which indicate the vacuum structures
of the scalars, we will assume they are changing in a certain range, such as $0.5<\tan\beta,~\tan\phi<10$, $0.5\leq \tan\phi\leq 10$, which are permitted by the constraints in Ref. \cite{top-seesaw-tc}.

\end{itemize}
\section{The charged Higgs pair productions at the LHC}
In this section, we discuss charged Higgs pair production processes $gg\to
H^+ H^-$, $q\bar q \to H^+ H^-$ ($q=u,d,s,c,b$), in top-seesaw assisted
TC models. In these processes, some
couplings such as $ H^\pm \bar f f'$ ($f,f'=~t,~b,~T,~B$) and $S H^+_L H^-_L$
($S=h^0, ~H^0$) etc., contain the model-dependent parameters so
that it may be viable to probe the new physics theory at future
collider experiments, via the effects of these parameters.

The cross sections of the charged Higgs pair production at the LHC comes
mainly from the gluon gluon fusion $gg\to H^+ H^-$, and quark pair annihilation processes
$q\bar q\to H^+ H^-$. 
At the LHC, the parton level cross sections for $pp\to H^+ H^-$
 are calculated at the leading order as
\begin{equation}
\hat{\sigma}(\hat s)
=\int_{\hat{t}_{min}}^{\hat{t}_{max}}\frac{1}{16\pi \hat{s}^2}
\overline{\Sigma}|M_{ren}|^{2}d\hat{t}\,,
\end{equation}
with
\begin{eqnarray}
\hat{t}_{max,min} =\frac{1}{2}\left\{ m_{p_1}^2 +m_{p_2}^2 -\hat{s}
\pm \sqrt{[\hat{s} -(m_{p_1}+m_{p_2})^2][\hat{s} -(m_{p_1}
-m_{p_2})^2]} \right\},
\end{eqnarray}
where $p_1$ and $p_2$ are the first and the second initial particles
in the parton level, respectively. For our case, they could be gluon
$g$ and quarks $u$, $d$, $c$, $s$, $b$.

 The total hadronic cross section for $pp \rightarrow H^+_L H^-_L+X$ can be obtained by folding the
subprocess cross section $\hat{\sigma}$ with the parton luminosity
\begin{equation}
\sigma(s)=\int_{\tau_0}^1 \!d\tau\, \frac{dL}{d\tau}\, \hat\sigma
(\hat s=s\tau) ,
\end{equation}
where $\tau_0=(m_{p_1}+m_{p_2})^2/s$, and $s$ is the $p p$
center-of-mass energy squared. $dL/d\tau$ is the parton luminosity
given by
\begin{equation}
\frac{dL}{d\tau}=\int^1_{\tau} \frac{dx}{x}[f^p_{p_1}(x,Q)
f^{p}_{p_2}(\tau/x,Q)+(p_1\leftrightarrow p_2)],\label{dis-pp}
\end{equation}
where $f^p_{p_1}$ and $f^p_{p_2}$ are the parton $p_1$ and $p_2$
distribution functions in a proton, respectively.
 In our numerical
calculation, the CTEQ6L~\cite{Ref:cteq6} parton distribution function is
used and take factorization scale $Q$ and the
renormalization scale $\mu_F$ as $Q=\mu_F = 2 m_H$. The loop
integrals are evaluated by the LoopTools
package~\cite{Ref:LoopTools}.


\subsection{The calculation of the the cross sections of the charged Higgs pair productions at The LHC}
\begin{figure}[t]
\begin{center}
\epsfig{file=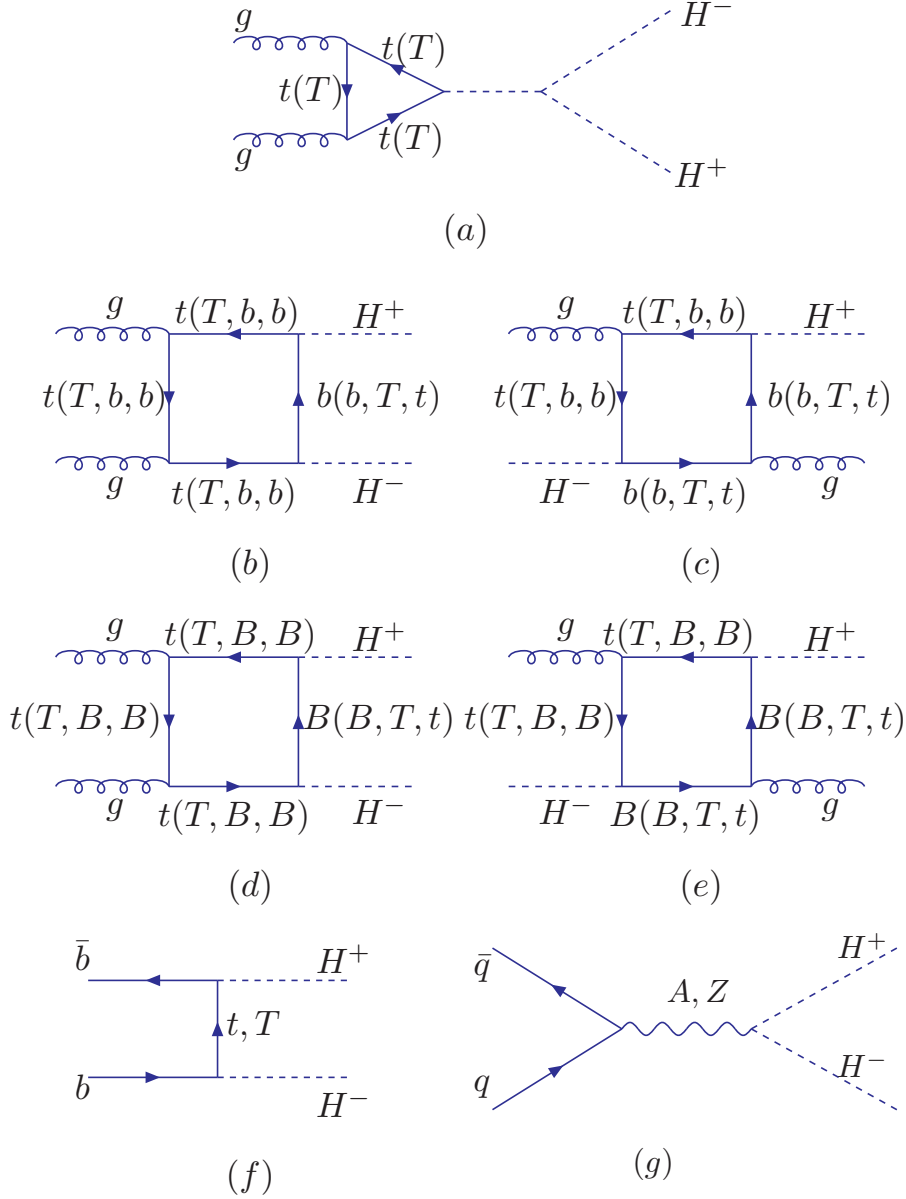,width=12cm}
\vspace{-0.4cm}\caption{Feynman diagrams for the charged Higgs pair production at the LHC via
gluon gluon fusion and the quark-anti-quark annihilation parton level processes
in the top-seesaw assisted TC models are demonstrated,
and $T,~B$ are the partner quarks of $t,~b$. Those obtained by exchanging
the two external gluon lines are not displayed here.}
\label{fig1}
\end{center}
\end{figure}
 In this section, we study cross sections for the
double charged Higgs production processes $gg\to H^+ H^- $,  $q\bar q \to  H^+ H^-$.
 Throughout this paper, we take $m_t =173 $ GeV
\cite{topmass}, $\alpha_s(m_Z) = 0.118 $ \cite{datagroup} and neglect bottom quark mass.

As for the parameters in the present model, we will consider the masses of
light Higgs to be $126$ GeV, and the masses of the vector-like particle $m_T=m_B=5$ TeV.
Other parameters involved in these processes are the followings: the charged Higgs masses,
the mixing of the scalars $\tan\beta,~\tan\phi$,
the dynamical generation quark masses $\Sigma_{U,D}$,
and the fraction of the ETC sector to the third quark masses, $\epsilon_{t,b}$ parameters.
In the following, we will take the light charged Higgs mass $m_{H_L}$ in the range $200 - 1000$ GeV,
while the mass of the heavy one mass varies from $1000$ to $5000$ GeV. Since the parameters
$\Sigma_{D}$ and $\epsilon_b$ are not affected the results largely, we will fix them as
$\Sigma_{D} =200 GeV$ and $\epsilon_b =0.1$.
For other parameters, the ranges can be taken as: $0.5<\tan\beta<10$, $0.5<\tan\phi<10$,
$200<\Sigma_U\leq 4000$ GeV and $0 \leq \epsilon_t\leq 0.1$.

The parton processes $gg\to  H^+ H^- $, $q\bar q \to  H^+ H^-$ ($H^\pm~ = ~H^\pm_L, ~H^\pm_H$) can be produced
at the LHC, with the Feynman diagrams shown in Fig.\ref{fig1}, which are realized by the gluon gluon fusion
 and quark-anti-quark annihilation, respectively, so we will firstly discuss the $gg$ fusion and the $q\bar q$ annihilation
processes, respectively, and then sum them together to obtain the
total contributions.

\subsubsection{The Process $gg\to  H^+_L H^-_L $ and $gg\to  H^+_H H^-_H $ }
Due to the interactions in Eq.(\ref{hff-couplings}) and Eq.(\ref{hchch-couplings}),
the charged Higgs pair production
processes can be realized by the triangle s-channel and the box t-(u-)channel at the LHC,
as shown in Fig.\ref{fig1}.

\def\figsubcap#1{\par\noindent\centering\footnotesize(#1)}
\begin{figure}[t]%
\begin{center}
\hspace{-0.6cm}
\parbox{8.05cm}{\epsfig{figure=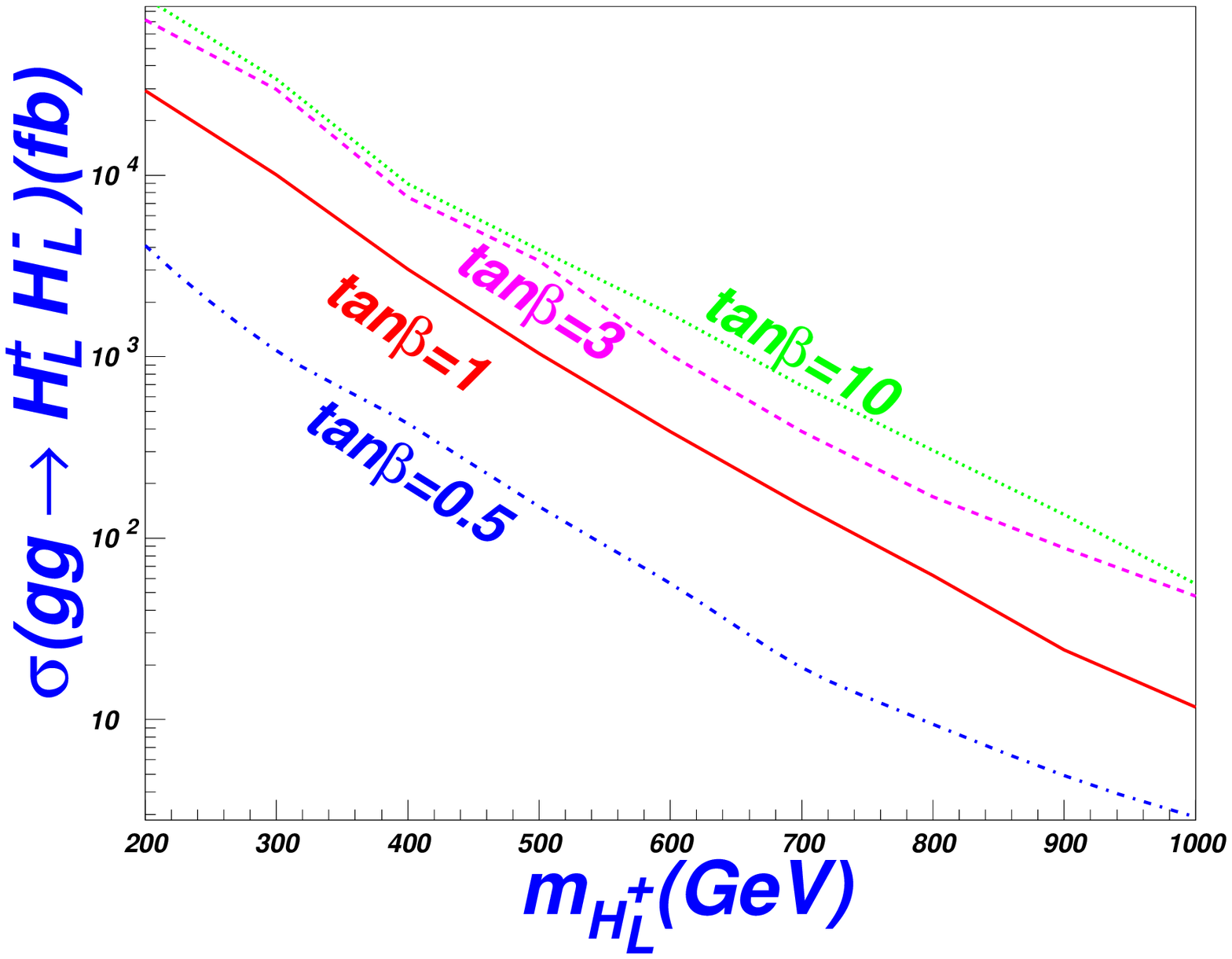,width=8.5cm} \vspace{-1.3cm} \figsubcap{a}}\hspace{0.1cm}
\parbox{8.05cm}{\epsfig{figure=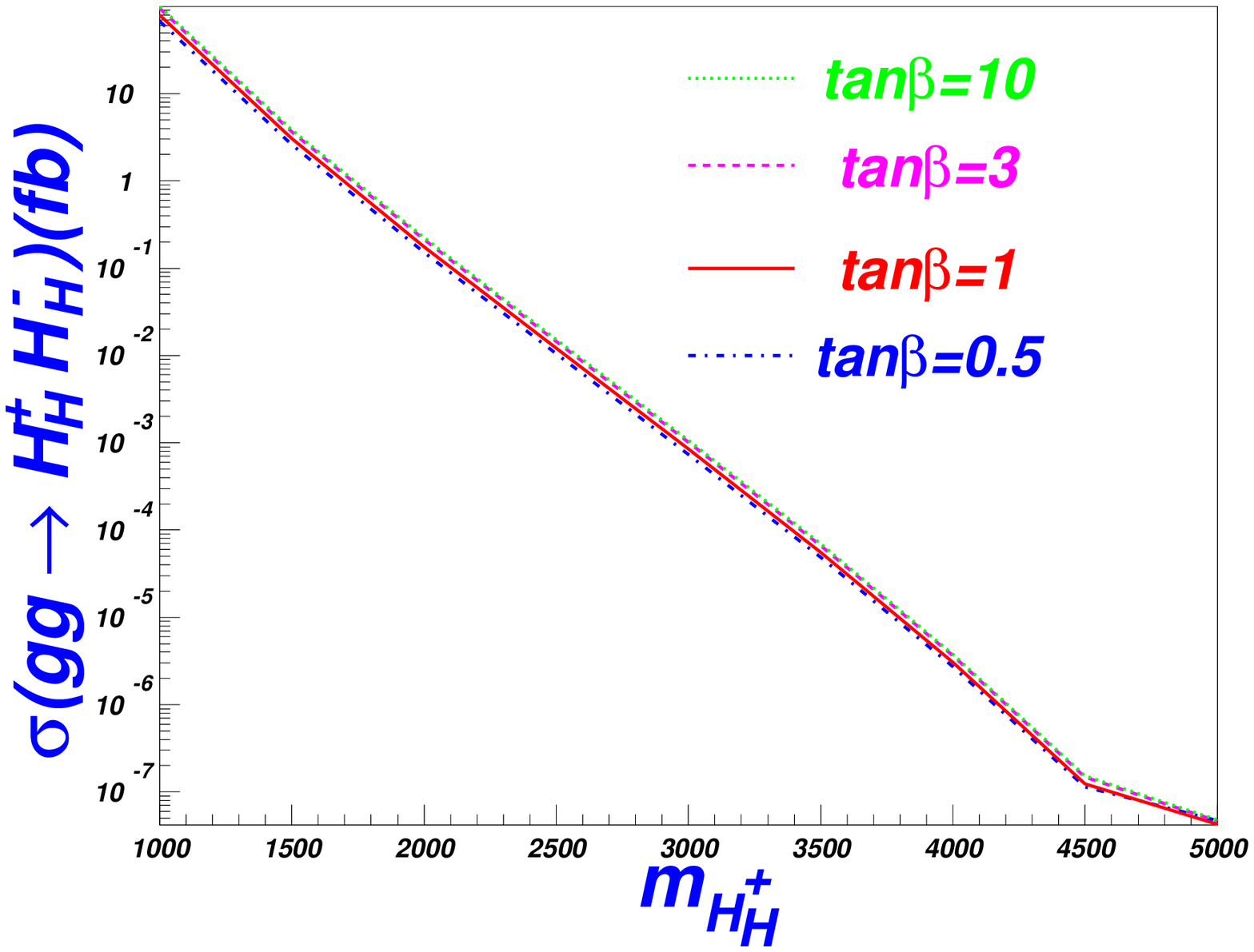,width=8.5cm} \vspace{-1.3cm}\figsubcap{b}}
\vspace{-0.4cm}\caption{ The cross section $\sigma$ of the processes $gg\to H^+_L H^-_L$ (a) and
$gg\to H^+_H H^-_H$ (b)
as a function of the charged scalar mass $m_{H}$ with
$\tan\beta=0,~1,~3,~10$ and $\sqrt{s}=14$ TeV .
 \label{fig2}  }
\end{center}
\end{figure}

\def\figsubcap#1{\par\noindent\centering\footnotesize(#1)}
\begin{figure}[t]%
\begin{center}
\hspace{-0.5cm}
\parbox{5.5cm}{\epsfig{figure=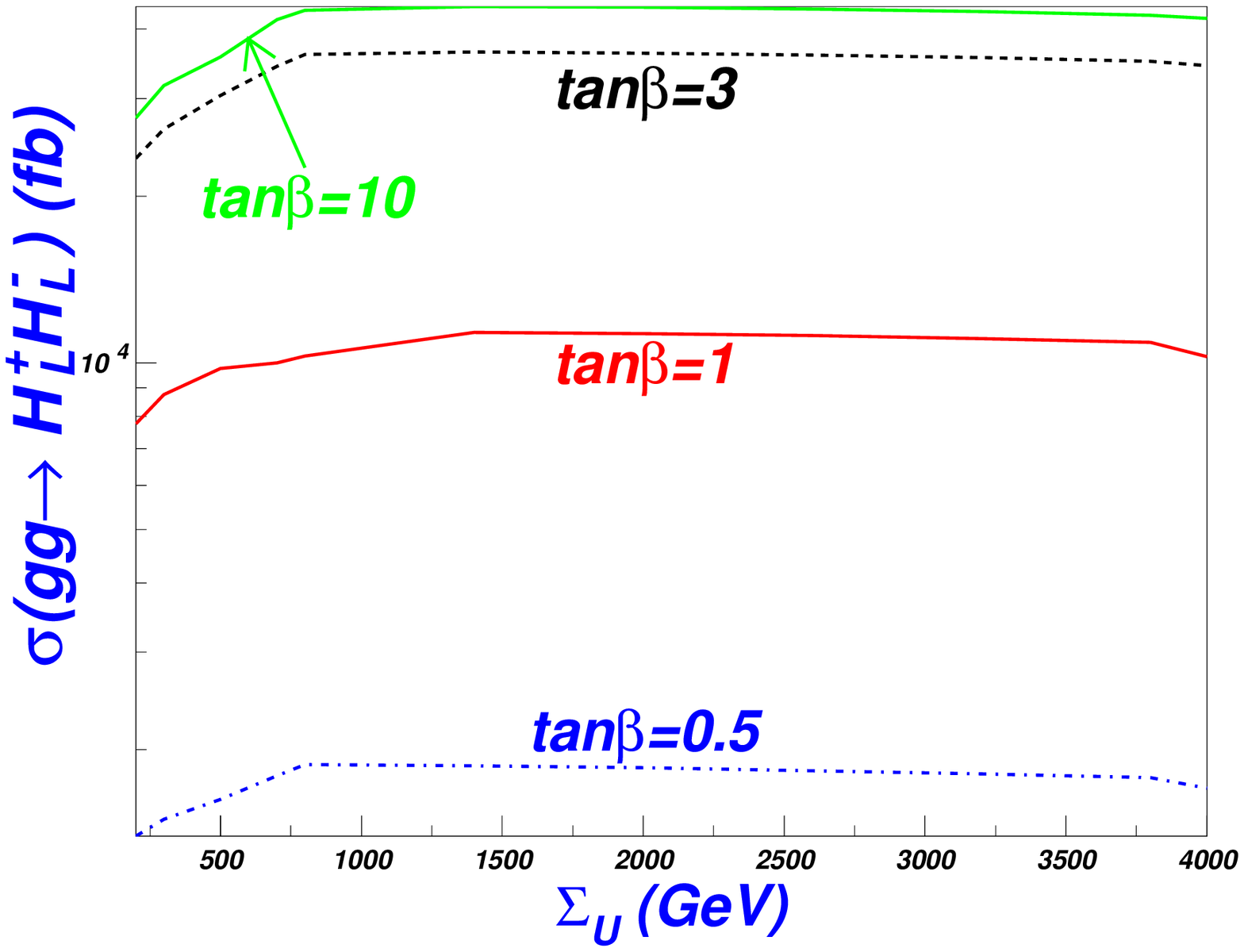,width=5.05cm} \vspace{-0.4cm} \figsubcap{a}}
\parbox{5.5cm}{\epsfig{figure=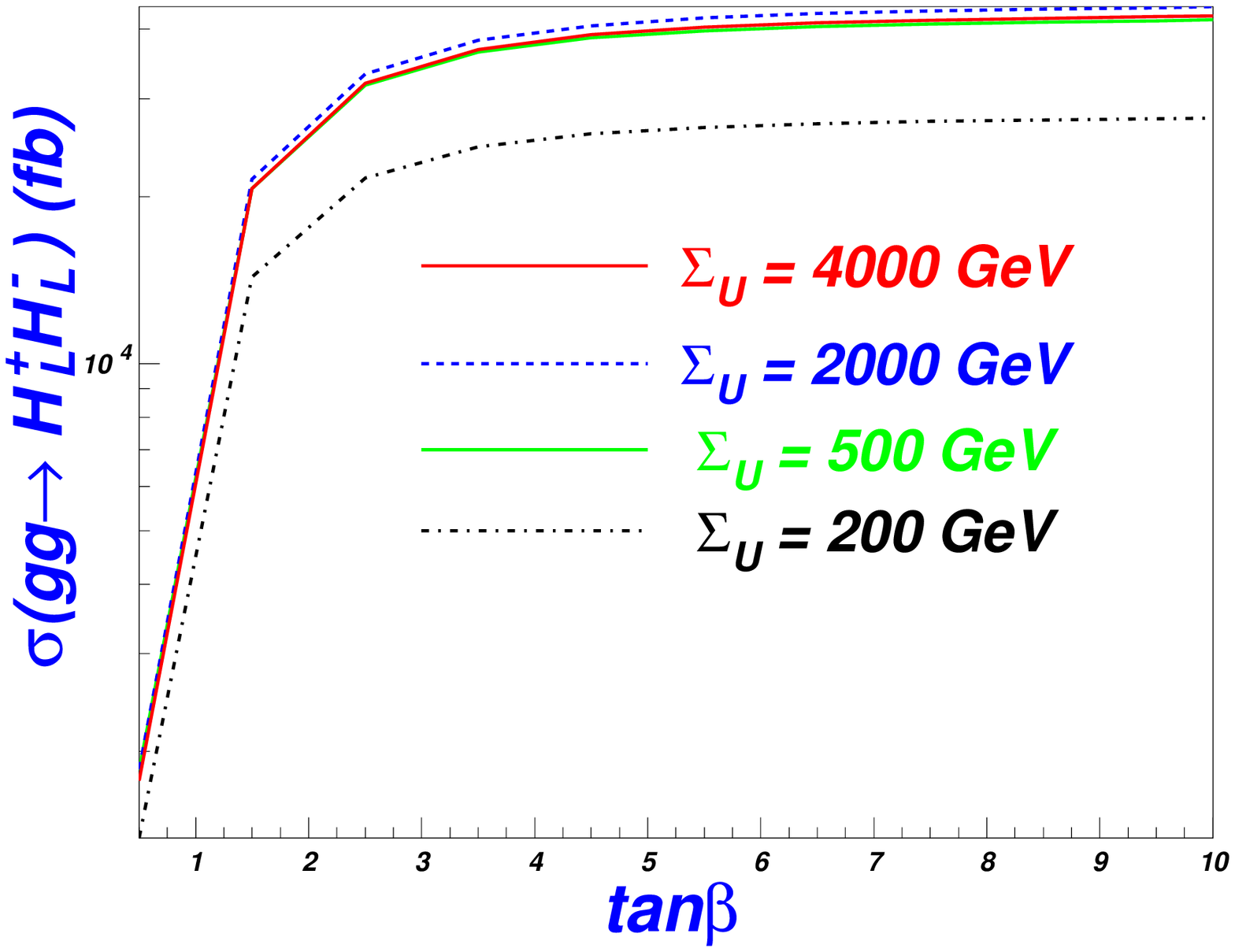,width=5.05cm} \vspace{-0.4cm}\figsubcap{b}}
\parbox{5.5cm}{\epsfig{figure=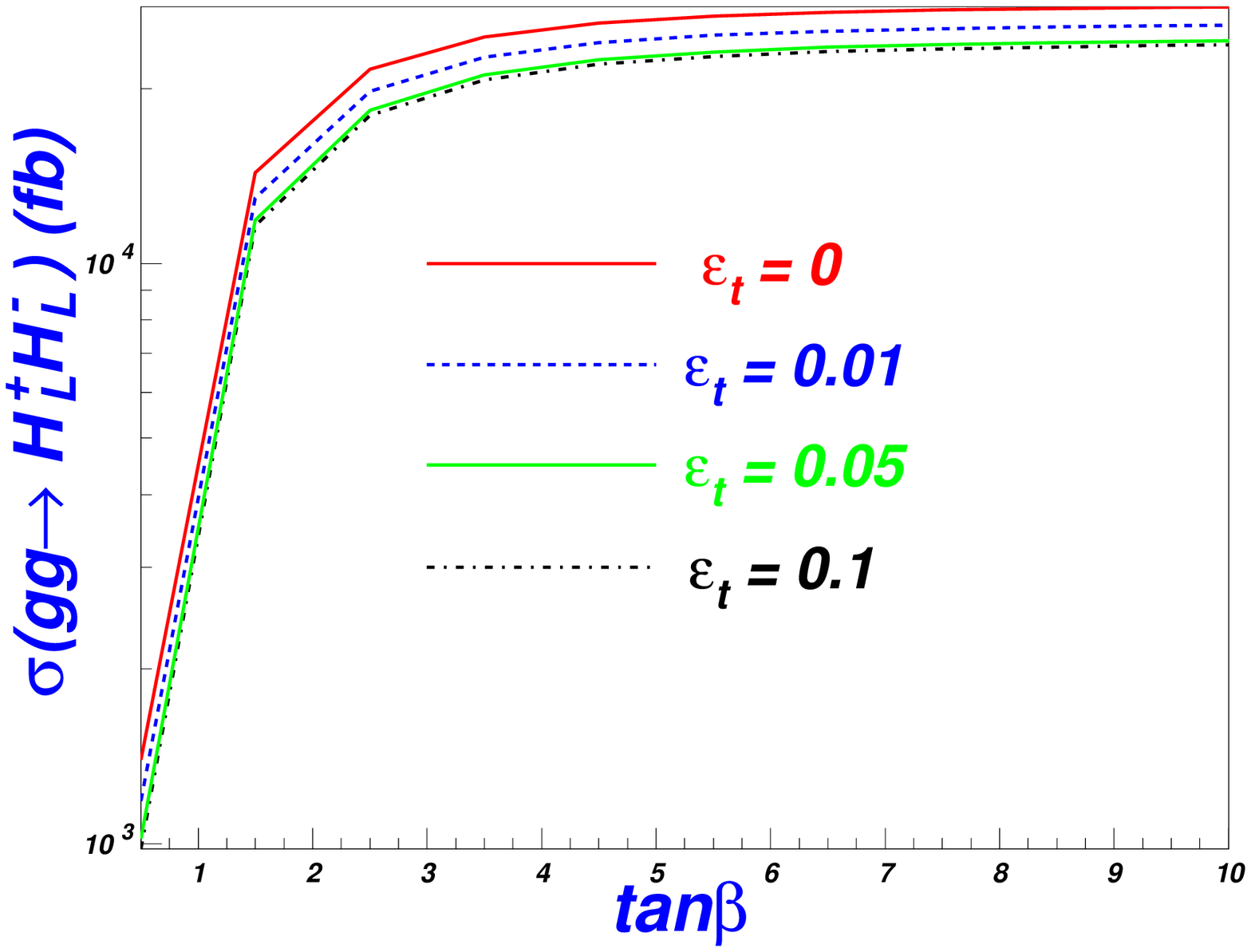,width=5.05cm} \vspace{-0.4cm}\figsubcap{c}}
\vspace{-0.4cm}\caption{The cross sections of the processes $gg\to H^+_L H^-_L$ are shown,
as a function of the dynamical fermion mass $\Sigma_U$ with $\epsilon_t =0.1$ and $\tan\beta=0.5,~1,~3,~10 $ (a), and of $\tan\beta$ with $\epsilon_t =0.1$ and
$\Sigma_U  =200,~500,~2000,~4000$ (b), and of $\tan\beta$ with $\Sigma_U  =200$ and
$\epsilon_t =0,~0.01,~0.05,~0.1$ (c), for $\sqrt{s}=14$ TeV, $m_{H_L}=300$GeV and $\tan\phi =3$.
 \label{fig3}  }
\end{center}
\end{figure}

\def\figsubcap#1{\par\noindent\centering\footnotesize(#1)}
\begin{figure}[t]%
\begin{center}
\hspace{-0.5cm}
\parbox{8.05cm}{\epsfig{figure=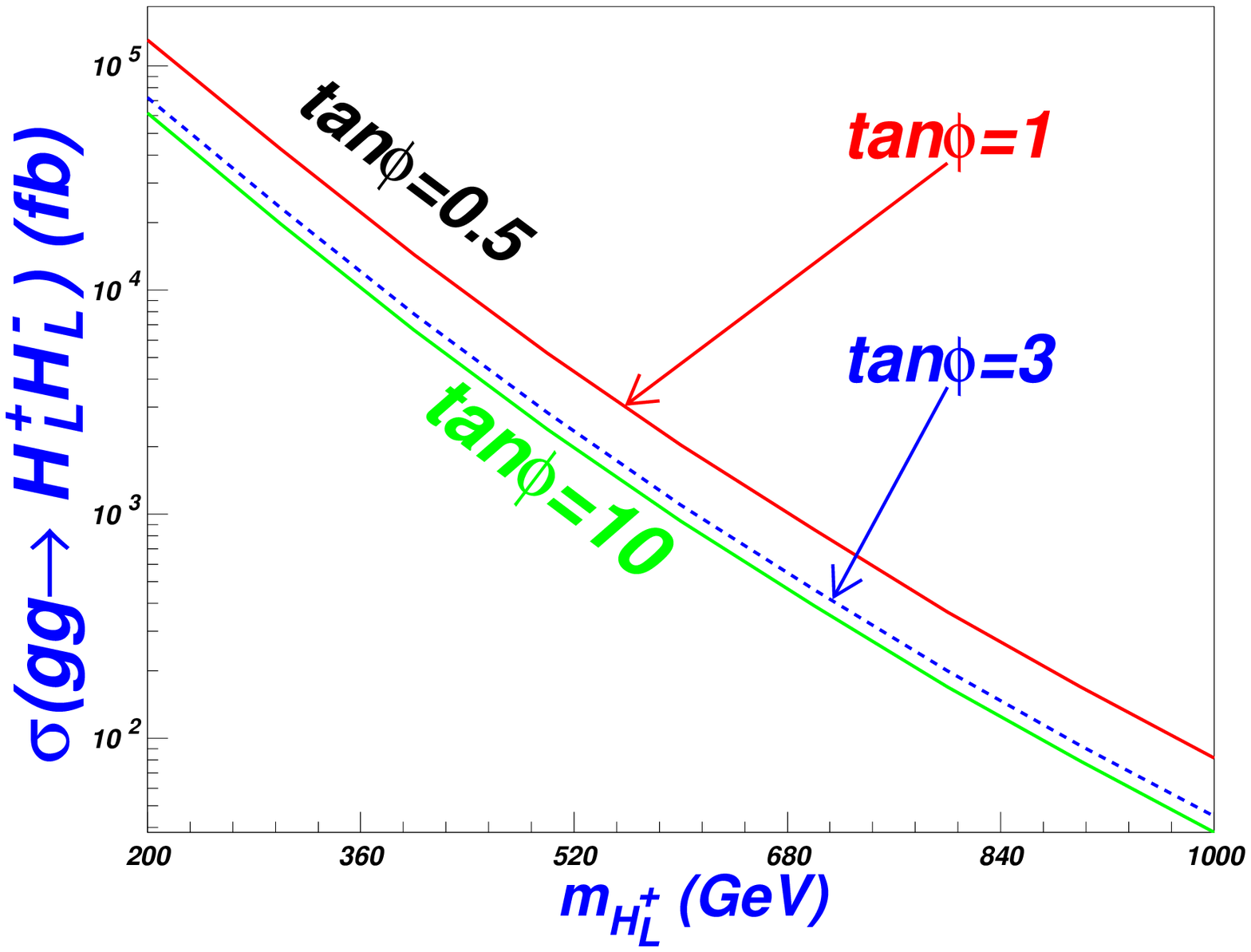,width=8.05cm} \vspace{-1.2cm} \figsubcap{a}}
\parbox{8.05cm}{\epsfig{figure=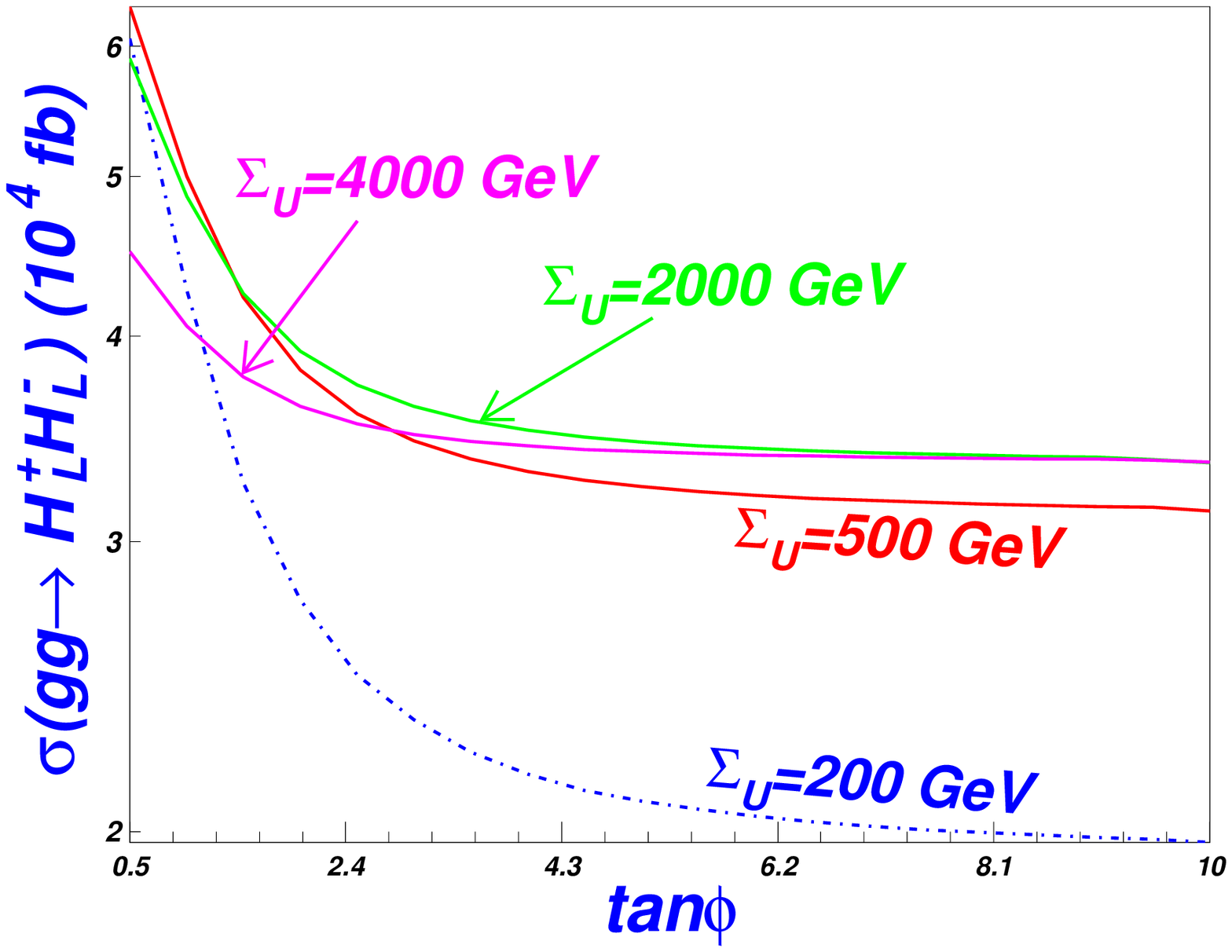,width=8.05cm} \vspace{-1.2cm}\figsubcap{b}}
\vspace{-0.4cm}\caption{The cross sections of the processes $gg\to H^+_L H^-_L$ are shown,
as a function of the charged scalar mass $m_{H_L}$ with
$\tan\phi=0.5,~1,~3,~10 $, $\tan\beta =3$ and $\Sigma_U =200$ GeV  (a), and of $\tan\phi$ with
$m_{H_L}=300$GeV, $\tan\beta =3$ and $\Sigma_U=200,~500,~1000,~2000,~4000$ GeV (b),
for $\sqrt{s}=14$ TeV,  and $\epsilon_t =0.1$.
 \label{fig4}  }
\end{center}
\end{figure}

The production cross sections  of the $ H^+_L H^-_L$ and $ H^+_H
H^-_H$ from the gluon gluon fusion are plotted in Fig.\ref{fig2}, for $\sqrt{s}=14$ TeV,
$\tan\phi=3$, $\epsilon_{t}=0.1$, $\Sigma_U=200$ GeV,
and $\tan\beta=0,~1,~3,~10$, as functions of the charged Higgs mass
$m_{H_L}$ and $m_{H_H}$, with light one changing from $200$ to $1000$ GeV,
and the heavy one from $1000$ to $5000$ GeV.

 From Fig.\ref{fig2}, we can see the cross section of this process $ H^+_L H^-_L$ is quite large,
which can arrive at $60$ pb in a favor parameter space, and in most of the
parameter space the cross sections can reach $1$ pb only if the charged Higgs is not too heavy.
 As was expected, the production rate decreases rapidly with the increasing charged Higgs mass since
the phase space are suppressed by the final particle masses, so it is natural that the process
$ H^+_H H^-_H$ is smaller than that of the former, about several fbs in
most of the parameter space. And with so heavy charged Higgs mass, the suppression was so strong that
the varying $\tan\beta$ values are not
affected the production rates at all, which shows clearly in Fig.\ref{fig2} (b).
In the following, we will only discuss the light charged Higgs pair production unless explicitly stated.

From Fig.\ref{fig2} (a) and Fig.\ref{fig3} (b),  we can also see the $\tan\beta$
dependence of the charged Higgs pair production processes
is strong, which is understandable, since $\tan\beta$ is closely connected to
the scalar VEVs  $v_1$ and $v_2$, and the in Fig.\ref{fig1} (a)-(e), the
dominant contributions are from the couplings $H^+ t(T) \bar b(\bar B) $
and the three scalars couplings $SH^+_L H^-_L$ ($S=h^0,~H^0$), which are
all related directly to the parameter $\tan\beta$.

In Fig.\ref{fig3} (a) we also show the cross sections as the functions of the $\Sigma_U$, which is the
dynamics fermion mass and find that the production rates are nearly a horizontal line with the varying
$\Sigma_U$. But in Fig.\ref{fig4} (b), for different $\Sigma_U$, the cross sections vary largely, especially, when
$\tan\phi$ is large. We can explain this as following: when other parameters contribute large,
that from $\Sigma_U$ is small, but with the increasing $\tan\phi$, the decreasing contributions from $\phi$
parameter, the effect of $\Sigma_U$ will stand out. The influence, however, is generally small.
 So in the following calculation, without affecting the results too much, we will take $\Sigma_U=200$ GeV.

As for the $\epsilon_t$ dependence, we show it in Fig.\ref{fig3} (c) and find the change of the cross sections
with the varying $\epsilon_t$ are quite limited, so we can conclude that $\epsilon_t=0.1$ is reasonable in
our computation and we will still take as that.

Just as that of the $\tan\beta$, we would like to know how the $\tan\phi$ affects the cross sections.
In Fig.\ref{fig4} (a) we give the cross sections varying as the light charged Higgs mass with
different $\tan\phi$, and just to find that the
effluence of changing $\tan\phi$ are quite small and the curves are almost the same, which is verified by Fig.\ref{fig4} (b),
from which we can see that when changing $\tan\phi$ from 0.5 to 10, the curves are almost coincided with each other, especially in the last part of them. Since $v_{TC}=v_{EW}\tan\phi$, we can conclude that the contributions of the TC section are small to the effective couplings $H^+t\bar b$ and $H^0H^+_L H^-_L$. Actually, this can be seen clearly from the couplings, for example, the terms closely connected with the $\tan\phi$ in couplings $H^+t\bar b$ can be write out explicitly $(-y_{TC}^bO_{23}+ y^t_{TC}O_{23} )+(-y_{TC}^bO_{23}- y^t_{TC}O_{23} )\gamma^5$, and the coefficients $y_{TC}\sim \epsilon_{t,b}<0.1$, which suppress the contributions; Moreover, the mixing $O_{23}$ decreases largely with the increasing $\tan\beta$, which also suppress the contributions largely.


From the discussion above we can see that the cross sections of the charged Higgs pair production
from the $gg$ fusion decline largely with the increasing charged Higgs masses,
while, at the same time, the parameters $\Sigma_U$, $\tan\beta$, $\epsilon_t$,  and $\tan\phi$
will also contribute to the production rates, which increase
with the increasing $\tan\beta$ and $ \Sigma_U$ (though very small), and decrease with the
increasing $\tan\phi$ and $\epsilon_t$.

We, in this paper, have found a very large rate for the $gg$ fusion production of pair charged Higgs bosons, which seems
in contrast with some existing results. For instance, in Ref. \cite{9707430},
for reasonably similar values of the parameters, the cross section of $gg\to H^+H^-$ is hardly
larger than a few fb with the same collider parameters. To explain this clearly, we write down the
couplings explicitly. In Ref. \cite{9707430}, the $H^-\bar t b$ coupling is $-\frac{1}{\sqrt{2}}[g_t\cot\beta+g_b\tan\beta +
(-g_t\cot\beta+g_b\tan\beta)\gamma^5]$ with $g_{t,b}=m_{t,b}/v_{EW}$, and to simplify the discussion, we neglect $g_b$ terms for small $m_b$ , then the coupling can be written as $-g_t/\sqrt{2}\cot\beta(1-\gamma^5)$, which is inversely proportional to $\tan\beta$.
While in our case, the coupling of the $H^-\bar t b$ is $(-c_L^tS_R^bO_{21}+s_R^tc_L^b O_{22})+(-c_L^tS_R^bO_{21}-s_R^tc_L^b O_{22})\gamma^5$ (dismissing the small parts from ETC).
and we find that, approximately, it can arrive at $(1+\gamma_5)$ level in a larger parameter space, since
two of the parameters $c_L^{t,b},S_R^{t,b}$ can be easily close to $1$ with large $m_{T,B}$. So the coupling of $H^-\bar t b$ is about $3$ times
larger than that from the that in Ref. \cite{9707430}, with $g_t=1/\sqrt{2}$ and $\tan\beta=1.5$. Since there are two
$H^-\bar t b$ vertexes to the processes, for the cross sections the contributions will be amplified by fourth power, that is,
$3^4=81$ times larger than that in Ref. \cite{9707430}. Not to say the large $\tan\beta$, the ratio will be larger(of course, the $g_b$ terms will be large for very large $\tan\beta$). Furthermore, we can see from Fig.\ref{fig1} that the particle spectrum have been added by the third partner particles $T,~B$, which have contributions of almost the same size as the top and the bottom quarks, so the amplitude will be crudely quadrupled and the cross sections will be amplified $16$ orders. Thus the total cross sections will be multiplied by a factor $81\times 16=1296$. Of course, this analysis is very crude only as a sketchy estimate.
Therefore we conclude that the cross sections in the top-seesaw assisted TC models may be much larger than those in some new physics models.

\subsubsection{$qq\to  H^+_L H^-_L $  }
\def\figsubcap#1{\par\noindent\centering\footnotesize(#1)}
\begin{figure}[t]%
\begin{center}
\hspace{-0.5cm}
\parbox{7.5cm}{\epsfig{figure=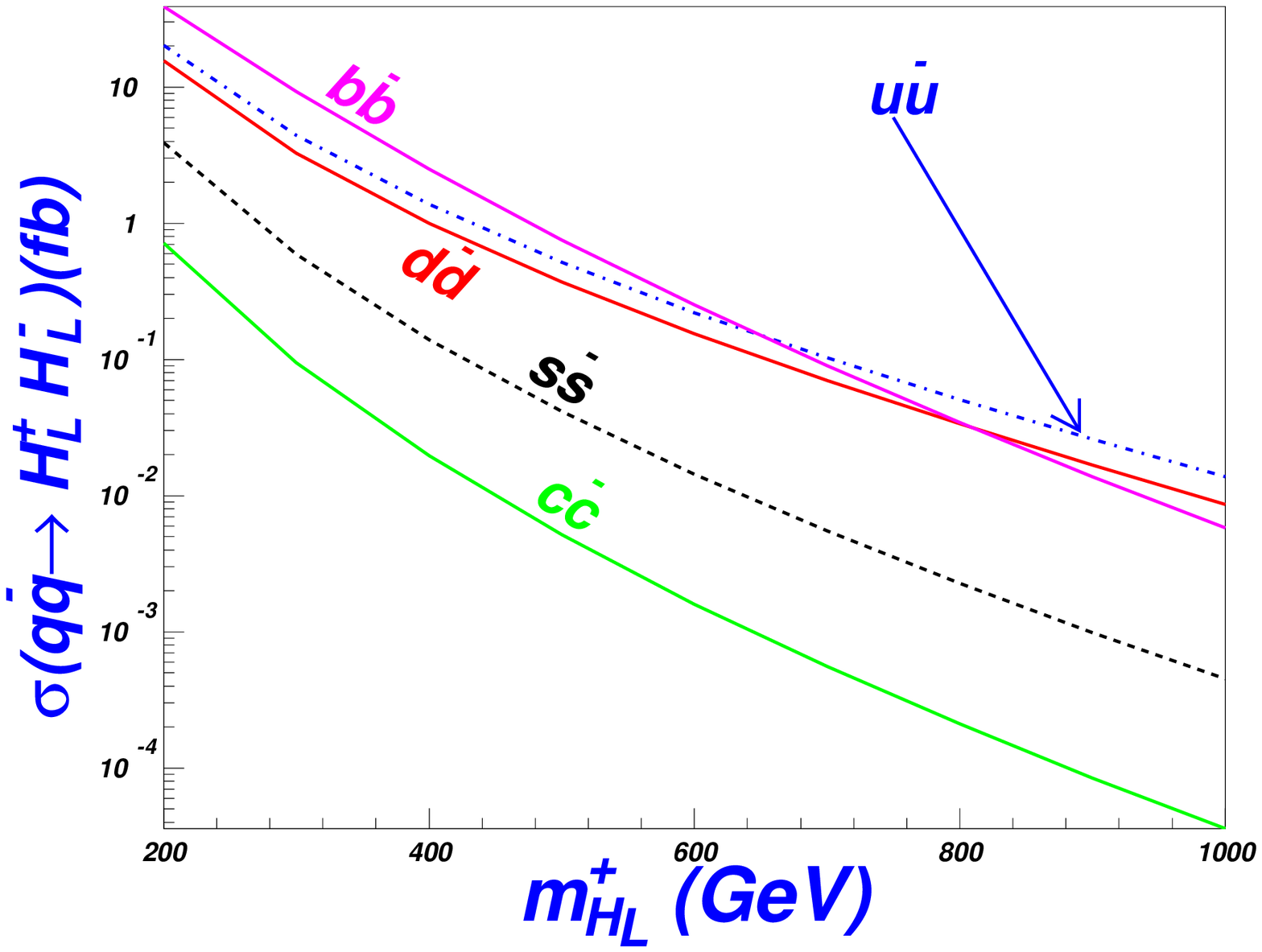,width=7.5cm} \vspace{-1.1cm} \figsubcap{a}}
\parbox{7.5cm}{\epsfig{figure=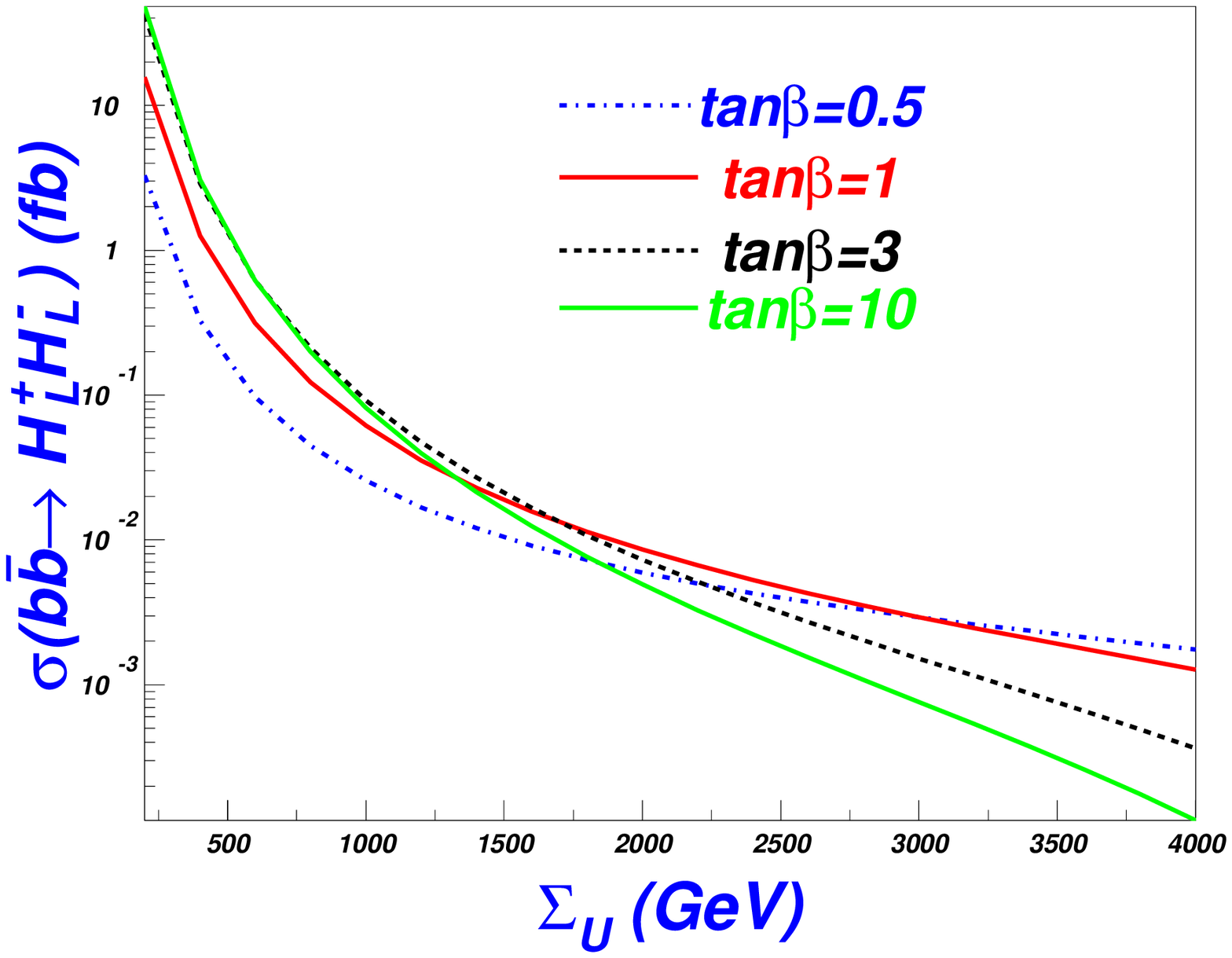,width=7.5cm} \vspace{-1.1cm}\figsubcap{b}}
\vspace{-0.4cm}\caption{ The cross section $\sigma$ of the processes $q\bar q\to H^+_L H^-_L$
($q=u~,d,~c,~s,~b$) as a function of the charged scalar mass $m_{ H_L}$ (a) and
that of the processes $b\bar b\to H^+_LH^-_L$
 as a function of the parameter $\epsilon_t$ (b), respectively, with
 $\sqrt{s}=14$ TeV are shown.
 \label{fig5}  }
\end{center}
\end{figure}
Here, the $ H^+_L H^-_L $ productions from different parton level have distinct cross sections
since the couplings and the parton distribution functions are different,
and there is not ony s-channel but also t-channel in $ b\bar b \to H^+_L H^-_L $ production, just as
shown in Fig.\ref{fig1}.

The s-channel processes such as Fig.\ref{fig1} (g), though the parton distribution functions could
be larger for the $u\bar u$ and $d\bar d$ initial state, may be
relatively small in view of the center-of-mass suppression effects.

At the same time, the t-channel coupling strengths may be larger than those of the
s-channel. In Fig.\ref{fig1} (f), For instance, the strengthen of $ H^+t\bar b \sim 1$, which
is larger than that of $Z H^+_L H^-_L $ and $\gamma H^+_L H^-_L $ (which are about $\sim e$)
 in the s-channel processes, so
the cross sections of the parton level processes
like $u\bar u(d\bar d, s \bar s) \to Z,\gamma \to  H^+_L H^-_L$ are smaller
than those of the others though with larger parton distribution functions. These can be seen clearly
in Fig.\ref{fig5}.

From Fig.\ref{fig5}, we can also see that, in most parameter space, the largest channel of the processes
$qq\to H^+_L H^-_L $ is the $b\bar b$ channel, which is easy to understand since, in
Fig.\ref{fig1}, the t-channel processes (f) are free of the center-of-mass depression
and the vertex of $ H^+t\bar b $, is in general, larger than that of $Z H^+_L H^-_L $ and $\gamma H^+_L H^-_L $, $\sim e$.

We also show the $\tan\beta$ and the $\Sigma_U$ dependence, respectively, of the
 cross sections from the $b\bar b$ annihilation for $m^+_{H_L} = 300$ GeV
in Fig.\ref{fig5} (b). We can see clearly that the production
rates decrease with increasing $\Sigma_U$, while for different $\tan\beta$, the production rates
 do not change much.

Comparing Fig.\ref{fig2}, Fig.\ref{fig3} and Fig.\ref{fig4} with Fig.\ref{fig5},
we can see that the contributions from gluon gluon
fusion is much more important than those from the quark-anti-quark annihilation, and the former can be
about $2-3$ order larger than the latter.

\subsubsection{The total contribution for the pair production of the light charged Higgs at the LHC}
Here we sum all the contributions, just as shown in Fig.\ref{fig6},
from which we can see the total pair production
rates of the light charged Higgs are related to the charged Higgs mass and
 the production probability with $\sqrt s = 14 $ TeV is larger than
$100$ fb for $m_H = 600 $ GeV in a large parameter space.
While, for the good case, for instance, for $m_H = 200 $ GeV,
the cross section can arrive at serval tens pb in most of parameter space.

\def\figsubcap#1{\par\noindent\centering\footnotesize(#1)}
\begin{figure}[t]%
\begin{center}
\hspace{-0.5cm}
\parbox{7.5cm}{\epsfig{figure=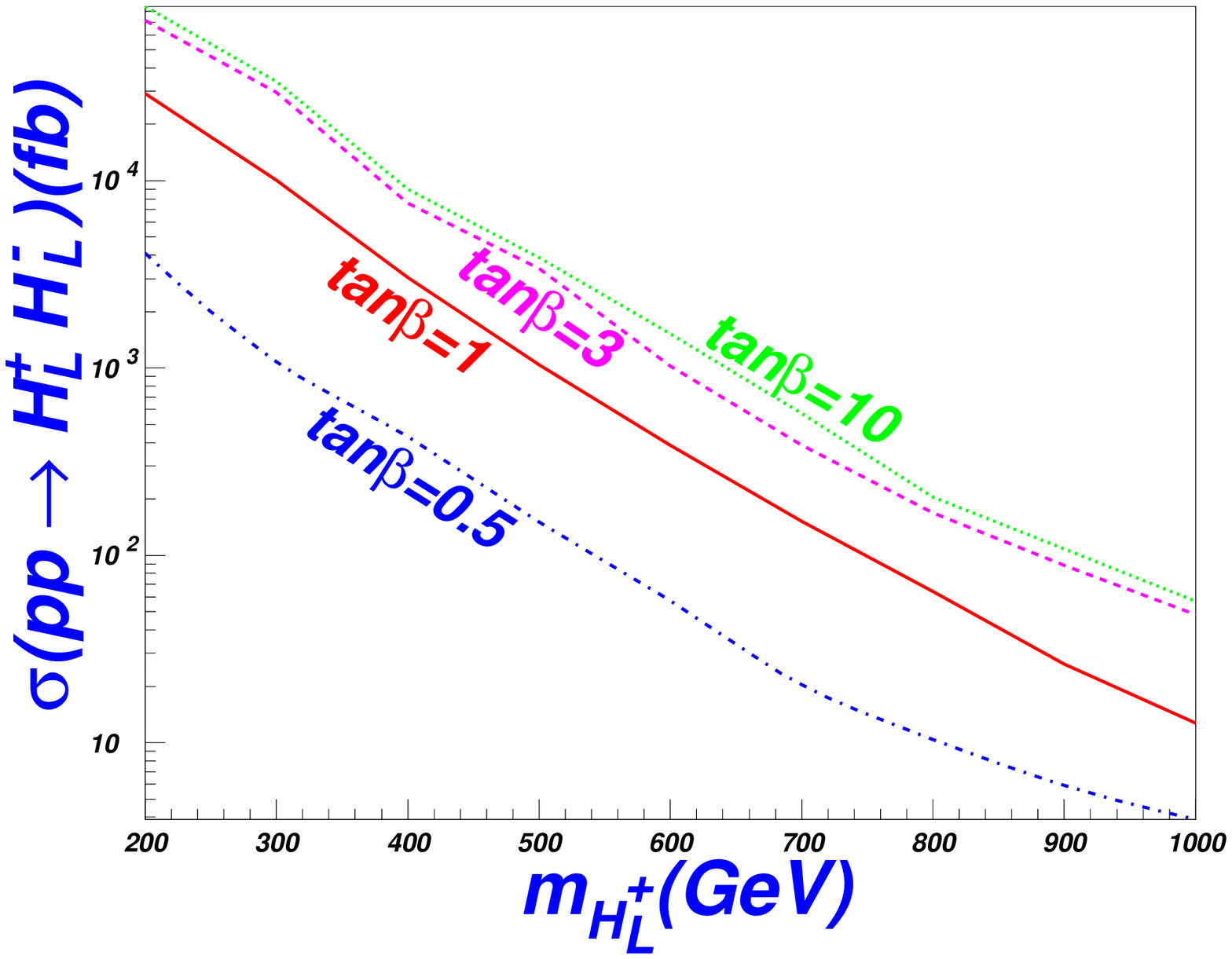,width=7.5cm} \vspace{-1.1cm} \figsubcap{a}}
\parbox{7.5cm}{\epsfig{figure=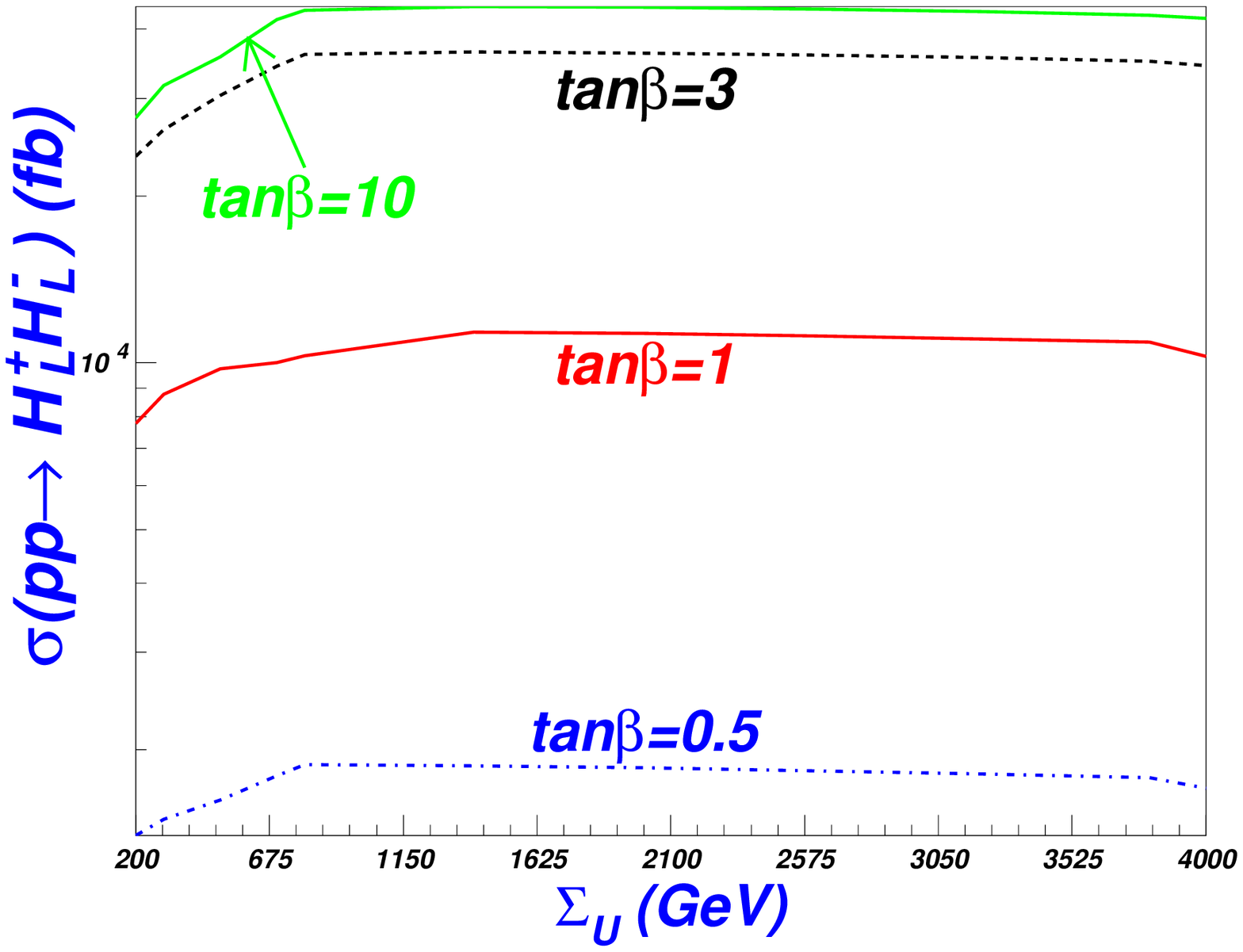,width=7.5cm} \vspace{-1.1cm}\figsubcap{b}}
\caption{ The total cross sections of the processes $pp\to H^+_L H^-_L$ are shown,
as a function of the light charged scalar mass $m_{ H_L}$  with $\Sigma_U=200$ GeV (a)
and of $\Sigma_U$ with $m_{ H_L}=300$ GeV (b) for $\sqrt{s}=14$ TeV, $\epsilon_t=0.1$.
 \label{fig6}  }
\end{center}
\end{figure}

From Fig.\ref{fig6} and Fig.\ref{fig3}, Fig.\ref{fig5}, we can see that both the
charged Higgs mass and the parameter $\tan\beta$ affects the production rates largely.
With different $\tan\beta$, the cross section may be $1$ even $2$ orders difference, which may
be used to constraint this parameter. For example, we can see from Fig.\ref{fig6} (b),
with the same parameters, when $\tan\beta=0.5$ GeV, the cross sections is about $2000$ fb,
while for  $\tan\beta=10$, the production rate increases to $40000$ fb when $m_{ H_L}=300$ GeV.

As for the effect of the parameter $\Sigma_U$, the influence is small comparing to that of $\tan\beta$, which
can be seen clearly in Fig.\ref{fig6} and Fig.\ref{fig3}.

\subsection{Backgrounds Analysis at the LHC}

For the light charged Higgs pair production $ H^+_L H^-_L$ at the LHC,
the charged light Higgs $ H^+$ decays mainly to $t\bar b$,
and top quark to $b$ quark,
charged lepton and the missing energy, i.e. the $4b+2l+\E_slash$
signal\footnote{Actually, usually only 2 bottom quarks are tagged,
so the signal is $2b+2l+2j+\E_slash$.} with $\E_slash$, the missing
energy, so the mainly SM backgrounds are $pp \to WWZjj$(with $Z$ to
$b\bar b$), $WWZZ$(with one $Z$ to $b\bar b$, the other to $jj$),
$WWhh$, $t\bar t W$ (with $W$ to two jets), $WWb\bar bjj$ and $t\bar
t jj$, where $h$ decays to $b\bar b$ and the $W\to l\E_slash$. Of
course, the signal cross sections would be reduced by the branching
ratios, $2/9\times 2/9\sim 0.05$ with $ Br(W->lv)=2/9$. 

The background production rates of the three processes, i.e,
$WWZjj$, $WWZZ$ and  $WWhh$ are quite small since there are more
than $3$ QED vertexes which suppress the strength. Considering the
branching ratio of $W$ and $Z$, the cross sections are at the level
of several tens of fb, so they are negligible in the SM background
discussion.  For $pp\to t\bar t W$, the production rate, about $500$
fb, similarly, the branching ratio of $W$ decaying to hadrons,
$2/3$, $t\to l\E_slash b$, $2/9$, then signal is about $4.6$ fb,
which is much smaller than that of the signal.

The process $pp\to WWbbjj$, is quite large, about $437$ pb,
multiplying by the $W$ branching ratios, $21$ pb. To suppress it, firstly, we
require the transverse momentum cut $p_T^j> 20$
Gev, since in the signal, the transverse momentum of the jets, which
are from the light charged Higgs, are large, while the transverse momentum of
the jets in the production $pp\to WWbbjj$, are much smaller. So the
background will be cut down largely, without losing signatures a lot
at the same time. Secondly, the light charged Higgs mass, or the top quark mass
reconstruction will be powerful to suppress the background since in
the signal the $Wb $ comes form the top quark, and the top quarks are from the charged Higgs,
  while in the background, it may not be the true case.

Another powerful background is $pp\to t\bar tjj$, about 227pb,
including the LO and the NLO contribution\cite{1002.4009}. The top
quark, however,  will decay to $Wb$ with $100\%$ percent, so the process
change into a part of the process $pp\to WWbbjj$ and it can also be
suppressed by the two methods mentioned above, i,e, the transverse momentum cut
and the mass reconstruction. 

From the discussion, we believe that the signal of the light charged Higgs will not be
reduced too much, 
while the background may be suppressed very much. Based on the
discussion above, we here arrive at the conclusion that the signal cross
sections arriving at $1000$ fb may be observable at the LHC.
Nevertheless, the discussion here is so crudely, and the precision
are far beyond control. We may, in the following work, debate the
observability at length.

To draw a very crudely conclusion, for an integrated luminosity $100$ fb$^{-1}$ at
the LHC, the charged scalar pair production cross sections of $1000$ fb may be
the lower limit of the observability.

\section{summary and conclusion  }
We considered the charged Higgs pair productions in the top-seesaw assisted TC model, proceeding
through $gg\to  H^+ H^-$, $q \bar{q} \to  H^+ H^-$, as a probe of the model.
Since the backgrounds may be effectively suppressed by the scalar mass
reconstruction, these processes can be used to probe the model.
We found that these charged Higgs pair productions in different collisions can
play complementary roles in probing the top-seesaw assisted TC model:

For the heavy charged Higgs pair production at the LHC, the cross section are quite small with the
increasing final particle masses, so we will not discuss little about that.

At the LHC, for the light charged scalars, the cross sections are large,
 and we have discussed the rates at the two parton level, i.e, the gluon gluon fusion and
quark-anti-quark annihilation, and compared their relative contribution. We find that the
contribution from the former is much larger than that from the latter.

After simple discussion of the backgrounds, for the $ H^+_L H^-_L$ production at the LHC, the
processes  may be detectable when the cross sections reach $1000$
fb, as discussed in the above section. For the process $gg\to  H^+_L H^-_L$,
the cross section can reach $1000$ fb in most of the parameter
spaces, which contributes large for this charged production.
 For $b\bar b\to  H^+_L H^-_L$ and $u\bar u\to
 H^+_L H^-_L$, the cross sections can arrive at several tens of fbs in most
of the parameter spaces, which are much smaller than those of the gluon gluon fusion, so
the main contribution is from the gluon gluon fusion.


As a conclusion, as long as the charged scalars are not too heavy, e.g.,
below $600$ GeV, the productions might be detectable at the LHC.
In general, the light charged Higgs pair productions have
larger possibility to be detected since their couplings to $t \bar
b$ are large.
We see from the figures listed above that in a large part of the
parameter space the cross sections of the scalar pair productions
can reach the possible detectable level, $1000$  fb for the LHC.
 Therefore, the pair productions of charged Higgs may
serve as a good probe of the top-seesaw assisted TC model.

\section*{Acknowledgments}
This work was supported by the Natural Science Foundation of China under the grant numbers 11105125 and 11105124 and the Excellent Youth Foundation of Zhengzhou U. under the grand number 1421317053.

\end{document}